\pdfoutput=1
\documentclass[8.5pt,twoside,twocolumn]{article}
\usepackage[super,sort&compress,comma]{natbib} 
\usepackage{times}

\usepackage{sectsty}
\usepackage{balance} 

\usepackage{bm}

\usepackage{graphicx} %eps figures can be used instead
\usepackage{amsmath,amssymb,amsfonts,mathrsfs}
\usepackage{epic,eepic}
\usepackage{graphics}
\usepackage{epsfig}
\usepackage{subfigure}
\newcommand{\de}{\partial}

\usepackage{bbm}
\newcommand{\id}{{\bm{\mathbbm{1}}}}

\newcommand{\ep}{\epsilon}

\newcommand{\be}{\begin{equation}}
\newcommand{\ee}{\end{equation}}
\newcommand{\bea}{\begin{eqnarray}}
\newcommand{\eea}{\end{eqnarray}}

\usepackage{amsmath}
\usepackage{amsfonts}
\usepackage{amssymb}
\usepackage{color}
\usepackage{graphicx}
\usepackage{breqn}

\usepackage{lastpage}
\usepackage[format=plain,justification=raggedright,singlelinecheck=false,font=small,labelfont=bf,labelsep=space]{caption} 
\usepackage{fancyhdr}
\begin{document}
\thispagestyle{plain}
\fancypagestyle{plain}{
%\fancyhead[L]{\includegraphics[height=8pt]{LH}}
%\fancyhead[C]{\hspace{-1cm}\includegraphics[height=20pt]{CH}}
%\fancyhead[R]{\includegraphics[height=10pt]{RH}\vspace{-0.2cm}}
\renewcommand{\headrulewidth}{1pt}}
\renewcommand{\thefootnote}{\fnsymbol{footnote}}
\renewcommand\footnoterule{\vspace*{1pt}% 
\hrule width 3.4in height 0.4pt \vspace*{5pt}} 
\setcounter{secnumdepth}{5}

\makeatletter 
\def\subsubsection{\@startsection{subsubsection}{3}{10pt}{-1.25ex plus -1ex minus -.1ex}{0ex plus 0ex}{\normalsize\bf}} 
\def\paragraph{\@startsection{paragraph}{4}{10pt}{-1.25ex plus -1ex minus -.1ex}{0ex plus 0ex}{\normalsize\textit}} 
\renewcommand\@biblabel[1]{#1}            
\renewcommand\@makefntext[1]% 
{\noindent\makebox[0pt][r]{\@thefnmark\,}#1}
\makeatother 
\renewcommand{\figurename}{\small{Fig.}~}
\sectionfont{\large}
\subsectionfont{\normalsize} 

\fancyfoot{}
%\fancyfoot[LO,RE]{\vspace{-7pt}\includegraphics[height=9pt]{LF}}
%\fancyfoot[CO]{\vspace{-7.2pt}\hspace{12.2cm}\includegraphics{RF}}
%\fancyfoot[CE]{\vspace{-7.5pt}\hspace{-13.5cm}\includegraphics{RF}}
\fancyfoot[RO]{\footnotesize{\sffamily{1--\pageref{LastPage} ~\textbar  \hspace{2pt}\thepage}}}
\fancyfoot[LE]{\footnotesize{\sffamily{\thepage~\textbar\hspace{3.45cm} 1--\pageref{LastPage}}}}
\fancyhead{}
\renewcommand{\headrulewidth}{1pt} 
\renewcommand{\footrulewidth}{1pt}
\setlength{\arrayrulewidth}{1pt}
\setlength{\columnsep}{6.5mm}
\setlength\bibsep{1pt}

\twocolumn[
  \begin{@twocolumnfalse}
\noindent\LARGE{\textbf{Mesoscale Structures at Complex Fluid-Fluid Interfaces: a Novel Lattice Boltzmann / Molecular Dynamics Coupling}}

\vspace{0.6cm}

\noindent\large{\textbf{Marcello Sega,$^{\ast}$\textit{$^{a,b}$} Mauro Sbragaglia,\textit{$^{b}$} Sofia S. Kantorovich\textit{$^{c,d}$} and Alexey O. Ivanov\textit{$^{d}$}} }\vspace{0.5cm}

%\noindent\textit{\small{\textbf{Received Xth XXXXXXXXXX 20XX, Accepted Xth XXXXXXXXX 20XX\newline
%First published on the web Xth XXXXXXXXXX 200X}}}

%\noindent \textbf{\small{DOI: 10.1039/b000000x}}
\vspace{0.6cm}
%Please do not change this text.

\noindent \normalsize{
Complex fluid-fluid interfaces featuring mesoscale structures with
adsorbed particles are key components of newly designed materials
which are continuously enriching the field of soft matter. Simulation
tools which are able to cope with the different scales characterizing
these systems are fundamental requirements for efficient theoretical
investigations.  In this paper we present a novel simulation method,
based on the approach of Ahlrichs and D\"unweg [Ahlrichs and D\"unweg,
\textit{Int. J. Mod. Phys. C}, 1998, \textbf{9}, 1429], that couples
the ``Shan-Chen'' multicomponent Lattice Boltzmann technique to
off-lattice molecular dynamics to simulate efficiently complex fluid-fluid interfaces.
 We demonstrate how this approach
can be used to study a wide class of challenging problems. Several
examples are given, with an accent on bicontinuous phases formation
in polyelectrolyte solutions and ferrofluid emulsions.  We also show
that the introduction of solvation free energies in the
particle-fluid interaction unveils the hidden, multiscale nature
of the particle-fluid coupling, allowing to treat symmetrically
(and interchangeably) the on-lattice and off-lattice components of
the system.}

\vspace{0.5cm}
 \end{@twocolumnfalse}
  ]

\section{Introduction}

\footnotetext{\textit{$^{a}$~Institut f\"ur Computergest\"utzte Biologische Chemie, University of Vienna, W\"ahringer Strasse 17, 1090 Vienna, Austria; e-mail marcello.sega@univie.ac.at}}
\footnotetext{\textit{$^{b}$~Department of Physics and INFN, University of Rome ``Tor Vergata'', Via della Ricerca Scientifica 1, 00133 Rome, Italy}}
\footnotetext{\textit{$^{c}$~Faculty of Physics, University of Vienna, Boltzmanngasse 5, 1090 Vienna, Austria}}
\footnotetext{\textit{$^{d}$~Institue of Mathematics and Computer Sciences, Ural Federal University, Lenin av. 51, Ekaterinburg, 620083, Russia}}

Mesoscale structures with colloidal suspensions and/or particles adsorbed at fluid-fluid interfaces are ubiquitous in nature and are a key component of many important technological fields \cite{Chaikin,Lyklema,Russel}. The dynamics of these particles, as well as that of polymers or polyelectrolytes that might be present in solution, lives on scales where thermal fluctuations and capillarity cannot be easily decoupled: the combined effect of electrostatic forces, surface tension, and liquid flow\cite{FanStriolo12,Staincik04,Chen13} governs the complex dynamics emerging during coalescence of Pickering emulsion droplets \cite{Pickering}; the effective magnetic permeability of ferrofluid emulsions \cite{Rosensweig} results from a delicate balance between droplets deformation/elongation and its magnetic moment, which may give rise to a non trivial dependence of the effective magnetic permeability in terms of the magnetic field \cite{IvanovKuznetsova12}; self-assembly at fluid-fluid interfaces, traditionally exploited in encapsulation, emulsification and oil recovery \cite{Lin03,Martinez08,Linchem}, has recently emerged in applications including functionalized nanomaterials with tunable optical, electrical or magnetic properties \cite{Collier97,Tao07,Cheng10} and still raises many challenges ahead; the conditions under which nanoparticles can adsorb to a fluid-fluid interface from suspension are still poorly understood and little is known on the microstructures forming at the interface \cite{garbin12}, also because thermal fluctuations compete with interfacial energy and may give rise to size–dependent self-assembly \cite{Lin03}.
This is an ideal test-bed for numerical simulations, as they can be used to characterize and investigate the influence of nano/microstructures, external perturbations (electric or magnetic fields, shear, etc.)  in ways that cannot be easily reproduced in laboratory experiments. In principle, atomistic molecular dynamics simulations could represent the most accurate microscopic approach, but the computational load restricts greatly their range of applicability unless large computational clusters are used \cite{kadau04,rapaport06,hess08,shaw09,klepeis09,loeffler12}.
A common solution is to employ mesoscale models from which hydrodynamics emerges spontaneously, therefore by-passing the need for interfacial treatment commonly required in other methods \cite{Prosperetti}, and  to couple them to a coarse-grained description of the solute or of  the particles in suspensions. The coarse-grained description allows to reduce the computational load by removing explicit solvent molecules while retaining the hydrodynamic interaction between other particles. 

Among the mesoscopic methods for the simulation of fluid dynamics, the dissipative particle dynamics \cite{hoogerbrugge92,espanol95},  the multiparticle collision  dynamics \cite{malevanets1999mesoscopic,ripoll2004low,kapral2008multiparticle} (also known as stochastic rotation dynamics \cite{ihle2001stochastic}) and the lattice-Boltzmann \cite{Benzi92} methods have been successfully employed to describe the dynamics of multicomponent or  multiphase fluids \cite{coveney97,moeendarbary09,kapral08,Benzi92,Chen98,Aidun10}. Lattice Boltzmann (LB), in particular, turned out to be a very effective method to  describe mesoscopic physical interactions and non ideal interfaces coupled to hydrodynamics \cite{Prosperetti} and many multiphase and multicomponent LB models have been developed, on the basis of different points of view, including the Gunstensen model \cite{Gunstensen,Gunstensen2}, the free-energy model \cite{YEO1,YEO2,YEO3} and the Shan-Chen model \cite{SC,SC2,SC5,ShanDoolen,ShanDoolen2}. Another different approach is that introduced by 
Melchionna and Marini Bettolo Marconi\cite{marconi09,marconi10,marconi11,marconi11b,melchionna11,melchionna12,marconi13}. The Shan-Chen model is widely used thanks to  its simplicity and efficiency in representing interactions between different species and different phases \cite{Kupershtokh,Hyv,Sbragaglia06,Sbragaglia09,CHEM09,Sbragaglia09b,Sbragaglia07,Shan06b,Sbragagliaetal12,VarnikSaga,JansenHarting11}.

Since the pioneering works by Ladd \cite{Ladd1,Ladd2}, the use of the LB method to study suspensions of solid particles attracted great interest in the LB community and several studies are now available \cite{Aidun,LaddVerberb,Lowe,Ding,Stratford,Joshi,JansenHarting11}, with applications ranging from biofluids to colloidal suspensions and emulsions \cite{Ramachandran06,Sun08,Stratford,Joshi}.  Some of the existing models also combine multiphase/multicomponent LB solvers with the Ladd (or closely related to) algorithm for suspended particles \cite{Stratford,Joshi,JansenHarting11}.  A different approach, explored first by Ahlrichs and D\"unweg \cite{ahlrichs98,ahlrichs99,ahlrichs01}, is based on an off-lattice representation of the solute, which is coupled to the LB fluid through a local version of the Langevin equation. Contrarily to the Ladd scheme, in this approach the particles can be penetrated by the fluid, but since they are off-lattice, a large variety of solutes can be easily modeled, allowing to represent structural details which are smaller than the lattice spacing, and to have a faster dynamics than the LB one. This approach has been successfully employed to describe polymer dynamics in confined geometries \cite{usta2005lattice}, polyelectrolyte electrophoresis \cite{grass2008importance,grass2009polyelectrolytes,grass2010mesoscale,grass2009optimizing}, colloidal electrophoresis \cite{lobaskin2004new,lobaskin2007electrophoresis,dunweg2008colloidal}, sedimentation \cite{kuusela2001velocity}, microswimmer dynamics \cite{lobaskin2008brownian}, biopolymers and DNA translocation \cite{fyta2008hydrodynamic,fyta2006multiscale}, DNA trapping \cite{kreft2008conformation}, thermophoresis \cite{hammack2011role} and electroosmosis \cite{smiatek2009mesoscopic}. 

Coupling off-lattice particles to one of the multicomponent LB methods would allow to address an even larger class of problems \cite{IvanovKuznetsova12,garbin12}. Rather remarkably, however, such a coupling has not been proposed so far. In order to fill this gap, in this paper we present a method that allows to model not only the mechanical effects of the particle-fluid coupling (through the Langevin friction), but also the solvation forces, in the context of a thermal Shan-Chen multicomponent fluid that satisfies the fluctuation-dissipation theorem (the latter requirement is of particular importance, as the characteristic energy scales in soft-matter systems are usually comparable with the thermal energy). We show how the method can be used to model several properties of particles interacting with interfaces, such as the particle contact angle or the interfacial tension reduction in presence of surfactants, and we apply the method to the problem of bicontinuous structure formation in presence of solvated polyelectrolytes and of droplet deformation in magnetic emulsion under the influence of an external magnetic field.

\section{Coupling the Shan-Chen multicomponent fluid to Molecular dynamics}

The fluctuating hydrodynamic equations that are simulated using the Shan-Chen approach \cite{SC,SC2,SC5,ShanDoolen,ShanDoolen2} are defined in terms of mass and momentum densities and the equations can be written as local conservation laws
\be\label{eq:NS}
\rho \left(\frac{\de}{\de t} {\bm{u}} + ({\bm{u}}\cdot {\bm{\nabla}})  {\bm{u}} \right)=-{\bm{\nabla}} p+{\bm{\nabla}} \cdot ({\bm{\Pi}}+\hat{{\bm{\sigma}}})+\sum_{\zeta} {\bm{g}}_{\zeta},
\ee
\be\label{eq:cont}
\frac{\de}{\de t} \rho_{\zeta}+{\bm{\nabla}} \cdot (\rho_{\zeta} {\bm{u}}) = {\bm{\nabla}} \cdot  ({\bm{D}}_{\zeta}+\hat{{\bm{\xi}}}_{\zeta}),
\ee
\be\label{eq:globalcont}
\partial_t \rho+{\bm{\nabla}} \cdot (\rho {\bm{u}}) = 0.
\ee
In the above equations,  the index $\zeta$ identifies different species, $\rho=\sum_{\zeta}\rho_\zeta$ is the total density and $p=\sum_{\zeta} p_{\zeta}=\sum_{\zeta} c_s^2 \rho_{\zeta}$ is the internal pressure of the mixture, where $c_s^2$ is the sound speed. The common baricentric velocity for the fluid mixture is denoted with ${\bm{u}}$. The diffusion current ${\bm{D}}_{\zeta}$ and the viscous stress tensor ${\bm{\Pi}}$, along with the associated transport coefficients and their relation to the fluctuating terms $\hat{{\bm{\sigma}}}$ and $\hat{{\bm{\xi}}}_{\zeta}$ are described in detail in the Appendix. The forces ${\bm{g}}_\zeta$ are specified by the following \cite{SC,SC2,SC5,ShanDoolen,ShanDoolen2} 
\begin{dmath}\label{eq:SCforce}
{\bm{g}}_{\zeta}({\bm{r}}) =  - \rho_{\zeta}({\bm{r}}) \sum_{{\bm{r}}'}\sum_{\zeta'}  g_{\zeta \zeta'} \rho_{\zeta'} ({\bm{r}}') ({\bm{r}}'-{\bm{r}})
\end{dmath}
where $g_{\zeta \zeta'}$ is a function that regulates the interactions between different pairs of components and ${\bm{r}}'$ a lattice site usually related to the lattice Boltzmann velocities, $({\bm{r}}'-{\bm{r}}) \propto w_i {\bm{c}}_i$, with $w_i$ suitable isotropy weights (see Eq. (\ref{isotropy}) in Appendix). For our purposes is important to note that Eq. (\ref{eq:SCforce}) can be approximated in the continuum by
\begin{equation}\label{Taylor}
{\bm{g}}_{\zeta}({\bm{r}}) \simeq-\rho_{\zeta}({\bm{r}}) \sum_{\zeta'} g_{\zeta \zeta'} {\bm{\nabla}} \rho_{\zeta'}({\bm{r}}).
\end{equation} 
At equilibrium, the model is characterized by a bulk free energy functional
\begin{dmath}
{\cal F}_{bulk}=\sum_{\zeta} c_s^2 \rho_{\zeta} \log \rho_{\zeta}+\frac{c_s^2}{2} \sum_{\zeta  \neq \zeta'} g_{\zeta \zeta'} \rho_{\zeta} \rho_{\zeta'}
\label{eq:fbulk}
\end{dmath}
which guarantees phase separation when the coupling strength parameter $g_{\zeta \zeta'}$ is large. With phase separation achieved the model can describe stable interfaces whose excess interfacial free energy can be approximated by the following \cite{SC,SC2,SC5,ShanDoolen,ShanDoolen2,CHEM09}
\begin{dmath}
{\cal F}_{int}=-\frac{c_s^4}{4} \sum_{\zeta \neq \zeta'} g_{\zeta \zeta'} {\bm{\nabla}} \rho_{\zeta} \cdot {\bm{\nabla}}\rho_{\zeta'}.
\label{eq:fint}
\end{dmath}

It is important to notice that in the Shan Chen approach the phase
separation emerges naturally thanks to the internal forces. The
interface is not imposed by external contraints and evolves
spontaneously according to Eqs.~(\ref{eq:NS}), (\ref{eq:cont}) and
(\ref{eq:globalcont}). Being the outcome of nearest neighbor sites
interaction, the interfacial region is diffuse and develops fully
over, typically, 8-10 lattice sites: a two-dimensional interface
(for a three-dimensional fluid) needs therefore to be defined using
an additional criterion such as, for example, the locus where the
two components have the same density.

The fluctuating hydrodynamics equations are solved by evolving in time the discretized probability
density $f_{\zeta i}({\bm{r}},t)$ to find at position ${\bm{r}}$ and
time $t$ a fluid particle of component $\zeta$ with velocity  ${\bm{c}}_i$ (here we are using the D3Q19 model with 19 velocities) according to the LB update scheme
\begin{equation}f_{\zeta i} ({\bm{r}} + \tau {\bm{c}}_i, t + \tau )= f_{\zeta i} ({\bm{r}},t) + \Delta_{\zeta i} +\Delta^{g}_{\zeta i} +\hat{\Delta}_{\zeta i}.\label{eq:LBMIX}
\end{equation} 
The term $\Delta_{\zeta i}$ represent the effect of collisions, while $\Delta^{g}_{\zeta i}$ and $\hat{\Delta}_{\zeta i}$ represent the effect of forcing and
thermal fluctuations, respectively. As a staring point for the development of the fluid-particle coupling, we implemented a fluctuating
Shan-Chen LB by extending the scheme proposed by D\"unweg, Schiller
and Ladd \cite{Dunweg,dunweg08,Dunweg2}, that uses the multi-relaxation time model (MRT) \cite{DHumieres02} and 
computes the evolution of $f_{\zeta i}$ in the space of hydrodynamic modes (see Appendix).

The coupling to off-lattice point particles is realized by evolving the position of the $i$-th particle ${\bm{r}}_i$ with a  Langevin-like equation of motion \cite{ahlrichs98}
\begin{equation}\label{NEWTON}
m{\bm{a}}_i = {\bm{F}} - \gamma \left[{\bm{v}}_i-{\bm{u}}({\bm{r}}_i)\right] + {\bm{R}},
\end{equation}
where, besides the conservative forces ${\bm{F}}$, a stochastic term
${\bm{R}}$ and a frictional force proportional to the peculiar
velocity (the particle velocity relative to the local fluid one,
${\bm{v}}_i-{\bm{u}}({\bm{r}}_i)$) are acting on the particle. The stochastic
term is a random force with zero mean and  variance related to the
friction coefficient $\gamma$ as $\left\langle R_a(t)R_b(t')
\right\rangle= 2 k_BT \gamma \delta(t-t')\delta_{ab}$.
This way, the Langevin-like equation acts as a local, momentum-preserving
thermostat, which guarantees that, at equilibrium, particles are
sampling the canonical ensemble\cite{dunweg08}.

Since the fluid velocities are computed only at grid nodes, the
velocity field at the particles position ${\bm{u}}({\bm {{r}}}_i)$
has to be interpolated, usually employing a linear scheme.  The
interpolation scheme is also used to transmit momentum back from
the particles to the fluid, in order to preserve linear momentum.
So far, the coupling scheme parallels that of Ahlrichs and D\"unweg
\cite{ahlrichs98}, but with this choice only one would fail to
embody the model with important physical features such as the
particles solvation free energy, which is fundamental to describe
the likelihood for a particle to be found in one or in the other
fluid component. 
In the remainder of this section we will introduce two new
particle-fluid forces, that constitute the core of the proposed
coupling scheme. This will extend the method of Ahlrichs and D\"unweg
to multicomponent fluids, with the original method becoming a
particular case of the new one. This choice has been made for the
sake of continuity, and will help, for example, comparing previous
simulation results obtained with the original single component
method and this novel one.  The MRT version of the three-dimensional
Shan Chen fluid here implemented is also, to the best of our
knowledge, introduced here for the first time, and we therefore
include the derivation of the algorithm in Appendix.

The effect of solvation forces can be introduced
in the continuum model (\ref{NEWTON}) by adding a term that is
compatible with the continuum approximation of the force (\ref{Taylor}),
i.e., by adding a force to model particle solvation,
${\bm{F}}^{\mathrm{ps}}$, that is proportional to the gradient of
the various fluid components, \begin{equation}\label{eq:solvation_kappa}
{\bm{F}}^{\mathrm{ps}}_i= -  \sum_{\zeta} \kappa_{\zeta} {\bm{\nabla}}
\rho_{\zeta}({\bm{r}}_i), \end{equation} and that drives particles
towards maxima ($\kappa<0$) or minima ($\kappa>0$) of each component.
The analogy of Eq.  (\ref{eq:solvation_kappa}) and (\ref{Taylor}) can
be made even more apparent by introducing a coarse-grained time
scale $\theta_t$ on which the fluctuating motion of particles is
fast with respect to the evolution of the hydrodynamic fields. In
this case, it is possible to compute the average force acting on
the particles at a given point in space ${\bm{r}}$
\begin{equation}\label{Force:ps} {\bm{F}}^{\mathrm{ps}}({\bm{r}})
=-  \sum_{i, \zeta}\kappa_{\zeta} \left\langle \delta
({\bm{r}}_i-{\bm{r}})\right\rangle_{\theta_t} {\bm{\nabla}}
\rho_{\zeta}({\bm{r}}_i), \end{equation} where
$\langle\ldots\rangle_{\theta_t}$ is a time average performed within
the coarse-grained time scale $\theta_t$. The analogy between the
latter force and the solvation one (\ref{Taylor}) is completed by
noticing that $\kappa_{\zeta}\sum_i \left\langle \delta({\bm{r}}-{\bm
{{r}}}_i)\right\rangle_{\theta_t}$ plays, formally, the role of
$g_{\zeta \zeta_p} \rho_{\zeta_p}({\bm{r}})$, therefore allowing
to describe the ensemble of particles as another fluid component
($\zeta_p$). This parallel makes however also clear, that the force
${\bm{F}}^{\mathrm{ps}}$ represents only half of what is needed  to
complete the analogy with Eq. (\ref{eq:SCforce}), since $\kappa_{\zeta}$
is equivalent to only one of the off-diagonal terms of $g_{\zeta
\zeta'}$, namely, the one responsible for the action of the fluid
component on the particles.

A symmetric term that models the action of the
particles on the fluid  (i.e., how the particles are solvating the
fluid) is in principle needed, and should consist of a force term
on the fluid nodes that depends on the gradient of the local particle
density. The LB fluid lives on lattice sites (${\bm {{r}}}$) while particles do not, i.e. ${\bm{r}}_i \neq {\bm{r}}$ due to
the continuum evolution (\ref{NEWTON}). In order to model the force
of the particles on the fluid, we take equation  (\ref{eq:SCforce}) and
specialize it to the fluid-particles link $({\bm{r}}_i-{\bm{r}})$.
We then consider all particles living in the cubic-lattice domains
sharing the common lattice-vertex site ${\bm{r}}$:
\begin{equation}\label{eq:solvation_lambda}
  {\bm{F}}_{\zeta}^{\mathrm{fs}}({\bm{r}}) = -\lambda_{\zeta} \rho_{\zeta}({\bm{r}}) \sum_{i, {\bm{r}}'} \Theta \left[\frac{({\bm{r}}_i-{\bm{r}})}{|{\bm{r}}_i-{\bm{r}}|} \cdot \frac{({\bm{r}}'-{\bm{r}})}{|{\bm{r}}'-{\bm{r}}|} \right]\frac{{\bm{r}}'-{\bm{r}}}{|{\bm{r}}'-{\bm{r}}|^2},
\end{equation}
with $\Theta(x)=1$ if $0<x<1$ and 0 otherwise. Again, the similarity with the fluid force equation (\ref{eq:SCforce}) is evident by identifying $\sum_i \lambda_{\zeta} \left\langle\Theta \left[\frac{({\bm{r}}_i-{\bm{r}})}{|{\bm{r}}_i-{\bm{r}}|}\cdot \frac{({\bm{r}}'-{\bm{r}})}{|{\bm{r}}'-{\bm{r}}|} \right]\right\rangle_{\theta_t}$ as equivalent to $g_{\zeta \zeta_p} \rho_{\zeta_p}({\bm{r}}')$.

While the force ${\bm{F}}_i^{\mathrm{ps}}$ acting on the particles has the clear effect of moving them towards regions of constant density, the consequences of ${\bm{F}}^{\mathrm{fs}}$ are less evident. If only one particle is present, the fluid nodes around it experience a force pointing towards the particle and therefore, depending on the sign of the coupling constant $\lambda_{\zeta}$, the fluid density will increase ($\lambda>0$) or decrease ($\lambda<0$) around the particle. The solvation force ${\bm{F}}^{\mathrm{fs}}$ can therefore 
be exploited to introduce an effective excluded volume (or solvation shell, for positive values of $\lambda$) for point particles.
We note that on the lattice, imposing $g_{AB}=g_{BA}$ is enough
to guarantee total momentum conservation,  because under this
condition the total force acting between every pair of nodes $\bm{r}$
and $\bm{r}'$ in Eq.\ref{eq:SCforce} is identically zero. With the
actual implementation of the coupling, however, the solvation forces
Eqs.~(\ref{eq:solvation_kappa}) and (\ref{eq:solvation_lambda})
alone can not guarantee momentum conservation as they have a different
functional form. For this reason the momentum gained by particles
due to the solvation forces (and vice versa) is transmitted back
to the fluid (to the particles) to conserve the total momentum, by
performing the same linear interpolation employed for the viscous
force~\cite{ahlrichs98}. 

This scheme is not the only one possible,
and instead of modelling the force between fluid and particles
starting from the continuum approximation, where the total momentum
gained by a particle is transferred back to the fluid, one could
implement a scheme where the momentum conservation is applied on a
per-node basis (therefore requiring only one type of solvation
force) thus making the analogy between particles and nodes even
deeper. We have decided however to implement the former scheme, for
the sake of continuity with the approach of Ahlrichs and D\"unweg,
and also because of the possibility of addressing a larger phenomenology
by being able of tune separately the action of fluid on particles
and vice versa (e.g. allowing to model the presence of excluded
volume independently from solvation forces), leaving the latter
approach for future investigations.

\section{Remapping to physical units}

A possible choice for reduced units is the one in which distances,
time intervals and energies are computed in units of the lattice
spacing $a$ and time interval $\tau$, and of the thermal energy
$k_BT$, respectively.  The electric charge is expressed also in
reduced units, and the strenght of the electrostatic interaction
is set by the Bjerrum length $\ell_B$, namely, the distance at which
two unitary charges interact with an energy which is equal to $k_BT$.
This choice of reduced units is employed thoroughout this work.

The limits of applicability of the LB method are set, at low Reynolds
numbers, by stability constraints (typically, $\eta/\rho>10^{-3}
a^2/\tau$ in order for the system to be able to dissipate
energy~\cite{Benzi92}) and by the requirement of fulfilling the
hydrodynamic limit of low Knudsen numbers $Kn = \eta / ( \rho c_s
L) \ll 1$. Here $L$ is the outer scale of the problem, typically,
the simulation box size. To make a practical example, in a simulation
with a box of edge $L \simeq 100 a$, a kinematic viscosity $\eta/\rho
\simeq 0.1 a^2/\tau$ satisfies both requirements. The choice of the
lattice spacing $a$ is also bound to the typical particle size
$R_p$. In order to avoid discretization effects, one should have $R_p\geq a$. If the particle represents
a monomer or a group of monomers in a polymer, then $a\simeq R_p
\simeq1$~nm. The value of the kinematic viscosity then sets the
time scale of the simulation: if the fluid is water, 
$\eta/\rho \simeq 10^{-6}$~m$^2$/s at room temperature, and 
$\tau\simeq 0.1$~ps. This represents a factor 100 with respect to
typical atomistic integration timesteps. If the particle represents,
instead, a colloid with $R_p\simeq 1\mu$m, this would imply that
$\tau\simeq 0.1 \mu$s. Note that in this way, to obtain a realistic
mapping to the viscosity of the fluid, we are renouncing to remap
correctly the speed of sound, which is bound to be $c_s=\sqrt{1/3}
a/\tau$, and therefore corresponding to about $5.8\times10^3$ and
$5.8$~m/s for the particles of radius 1~nm and 1~$\mu$m, respectively:
care has to be taken not to generate supersonic motion of the
particle in out-of-equilibrium simulations, which would compromise
the qualitative behavior of its dynamics. 
In the fluid-particle coupling, however, there is another constraint,
which comes from the stability of the molecular dynamics integration 
scheme. In ordere to integrate properly the Langevin equation, the
product of friction coefficient and integration timestep has to be
$\gamma \Delta t / m_p \ll 1 $ (although this limit can be
extended~\cite{vangunsteren82}), where $m_p$ is the mass of the
particle in molecular dynamics.  For large enough particles, Stokes
law $\gamma = 6 \pi \eta R_p$ links the hydrodynamic radius of the
particle to friction coefficient and viscosity, so that, with our
choices $\eta/\rho \simeq 0.1 a^2/\tau$ and $R_p\simeq a$, the
condition on $\gamma \Delta t/m_p$ becomes $\rho\Delta t \ll m_p
\tau / a^3$.  The choice of the integration timestep, which is
usually in the range $\Delta t \simeq 0.01 - 1 \tau$ (notice that
this is in lattice units) and of the particle mass then sets a limit
on $\rho$. In the common case of particles with a density not much
different from the solvent, the requirement becomes $\Delta t \ll
\tau$.

\begin{figure}
\begin{centering}
\includegraphics[width=1.0\columnwidth]{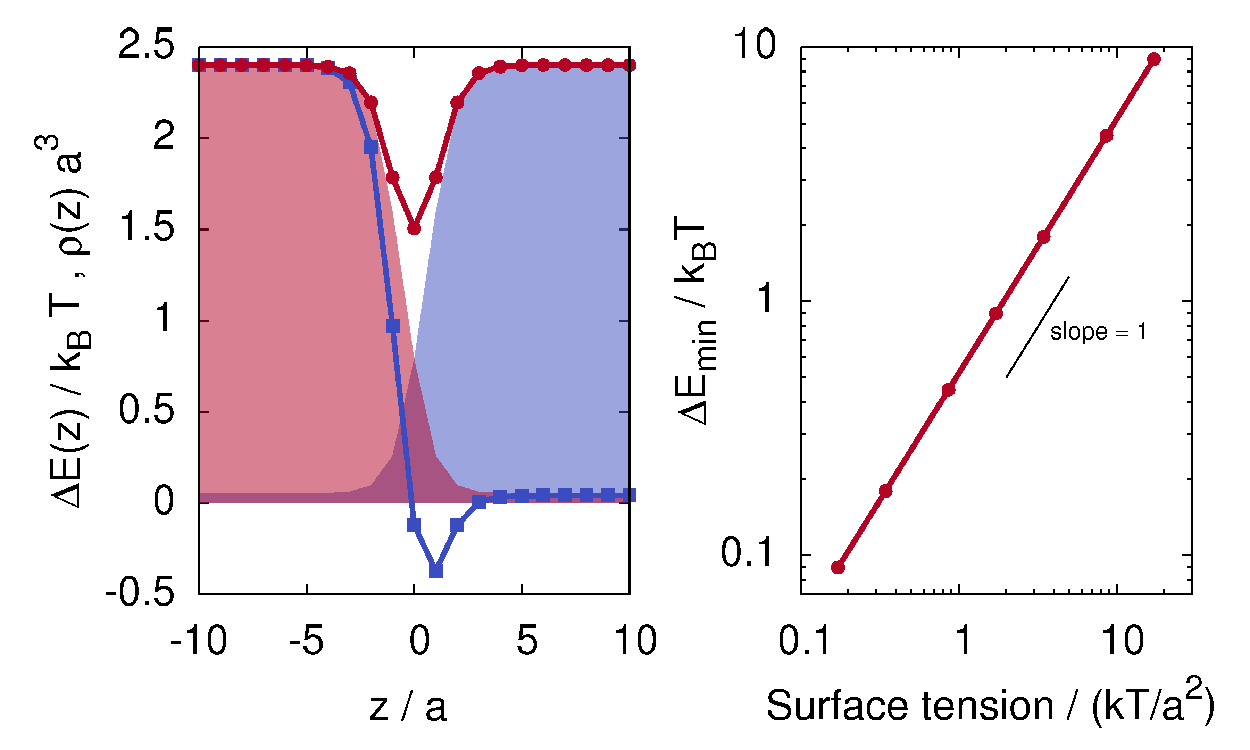}
\end{centering}
\caption{
Left panel: free energy profile $\Delta E(z)$ of a particle with
$\kappa_A=-k_BT a^3$, $\kappa_B=-2 k_BT a^3$ (squares). The free energy profile
is computed by integrating the force needed to keep the particle fixed, up to an immaterial constant. The
shaded areas show the fluid density profile $\rho(z)$: the expected
value of free energy difference between the two bulk regions is in
this case $\Delta E/k_BT=\Delta\rho_A a^3$. The free energy profile of a
particle with $\kappa_A=\kappa_B=-k_BT a^3$ is also reported
(circles).  Right panel: the depth of the free energy minimum of a
particle with $\kappa_A=\kappa_B=-k_BT a^3$ as a function of the
fluid surface tension. \label{fig:remapping}
} \end{figure}

Thermal fluctuations also have an important influence on the density.
The relative fluctuations of the populations define the Boltzmann
number $Bo=\sqrt{\langle f^2_i \rangle - \langle f_i\rangle^2}/
\langle f_i \rangle = \sqrt{k_B T / ( \rho c_s^2 a^3)}$: a value of
$Bo\simeq1$ will lead to negative populations, therefore, with
increasing temperatures the stability of the algorithm can be reached
by increasing the value of $\rho$. This condition is related to the limit for
an incompressible fluid, or low Mach numbers, which is in fact satisfied
when $\langle u^2 \rangle/c_s^2\ll1$, or, $\rho a^3\gg k_BT / c_s^2$. As a consequence, a lower limit for the surface tension that can be achieved
at a given temperature is set. In a Shan Chen fluid, rescaling $\rho \to
\alpha \rho$ and concurrently $g_{\xi\xi'}\to g_{\xi\xi'}/\alpha$
will allow to retain the mixing properties, so that the density profiles
$\rho(x)$ will keep the same shape, but also the surface
tension will increase by the factor $\alpha$ (see
Fig.~\ref{fig:remapping}).

The interpretation of the $\kappa_\sigma$ parameters in terms of
solvation free energies -- the quantitative control of which
guarantees that important properties like the partition coefficient
are properly modelled -- is easily recovered by noticing that in a demixing fluid with two components A and B, the 
work done to move a particle from the A-rich to the B-rich
region is \begin{equation} \Delta
E = -\sum_\xi \kappa_\xi \int \nabla \rho_\xi (\bm{r}) d\bm{r} =
-\sum_\xi \kappa_\xi \Delta \rho_\xi. \label{eq:solvation_density} \end{equation} Here $\Delta\rho_\xi$
is the density difference between rich and poor regions of component
$\xi$, so that $\kappa_\xi\Delta\rho_\xi$ is the solvation free energy
(in units of $k_BT$) of the particle in the fluid component $\xi$. Notice that if $\kappa_A=\kappa_B$, then the free energy profile is proportional to the total fluid density.
The free energy profile $\Delta E(z)$ of a particle moved across a planar interface 
is shown in Fig.~\ref{fig:remapping}, together with the density
profiles $\rho_A(z)$ and $\rho_B(z)$. The free energy profile is computed by integrating the force needed to keep the particle fixed. An (arbitary) offset has been added to the profile to match the numerical value of $\rho_A$ in the bulk phase. Given the choice of the
paramters, $\kappa_A=-k_BT a^3$ and $\kappa_B=-2k_BT a^3$, the
expected free energy difference betweeen the two bulk phases is
equal to the density difference of one phase across the interface
$\Delta E/k_BT =  \Delta \rho_A a^3$.

Even if the present method describes pointlike particles in presence
of diffuse interfaces, it is instructive to compare its results to
a simple but widely used mean-field model (see, e.g.,
Ref.~\cite{Pieranski80}) for spherical colloids and sharp interfaces.
In this model, the free energy profile of a colloid $c$ of radius
$R$, as a function of the distance $z$ from the interface, is written
in terms of the colloid-fluid surface energies $\gamma_{c\xi}$ and
fluid interfacial energy $\gamma_{AB}$ as $E(z/R) = \gamma_{cA}
2\pi R^2(1-z/R) +  \gamma_{cB} 2\pi R^2(1+z/R) - \gamma_{AB} \pi
R^2 (1-z^2/R^2)$.  The first two contributions are linear in $z$
as they are proportional to the fraction of the colloid surface in
contact with the fluid, while  the last contribution originates
from the missing A/B interface and is quadratic. The quadratic term
is responsible for the presence of an energy minimum located close
to the interface, also in case of equal wettability of the particle
with respect to the two fluid components. Despite the opposite
assumptions in the model and in the present simulation approach
(large particles and sharp interface in contrast to pointlike
particles and diffuse interface, respectively) it is interesting
to notice that since the total density of the fluid has a minimum
at the interface, the choice of positive solvation free energies
($\kappa_\xi>0$) for both components can lead to the appearance of
a minimum of the free energy at the interface.  The depth $\Delta
E_{\mathrm{min}}$ of the minimum, moreover, shows the same qualitative
dependence from the interfacial tension as in the model, $\Delta
E_{\mathrm{min}}\propto\gamma_{AB}$. This is shown in the right
panel of Fig.~\ref{fig:remapping} for systems with $\kappa_A=\kappa_B=k_BT
a^3$ and different surface tensions, obtained by keeping the product
$\Delta\rho_\xi g_{\xi\xi'}$ fixed.

Regarding the solvation force~Eq.(\ref{eq:solvation_lambda}), if
the magnitude of $\lambda_\xi$  is small enough not to perturb
significantly the fluid density, the particle free energy profile
is the same (modulo a factor of 1/2) as that obtained using
~Eq.(\ref{eq:solvation_kappa}) and the same numerical values for
$\kappa_\xi$. With growing values of $\lambda_\xi$, however, the
density profile is so much changed that it become possible to realize
a separation between the fluid and the particles.

\begin{figure}
\begin{centering}
\includegraphics[bb=135 280 451 582,clip,angle=90,width=0.39\columnwidth]{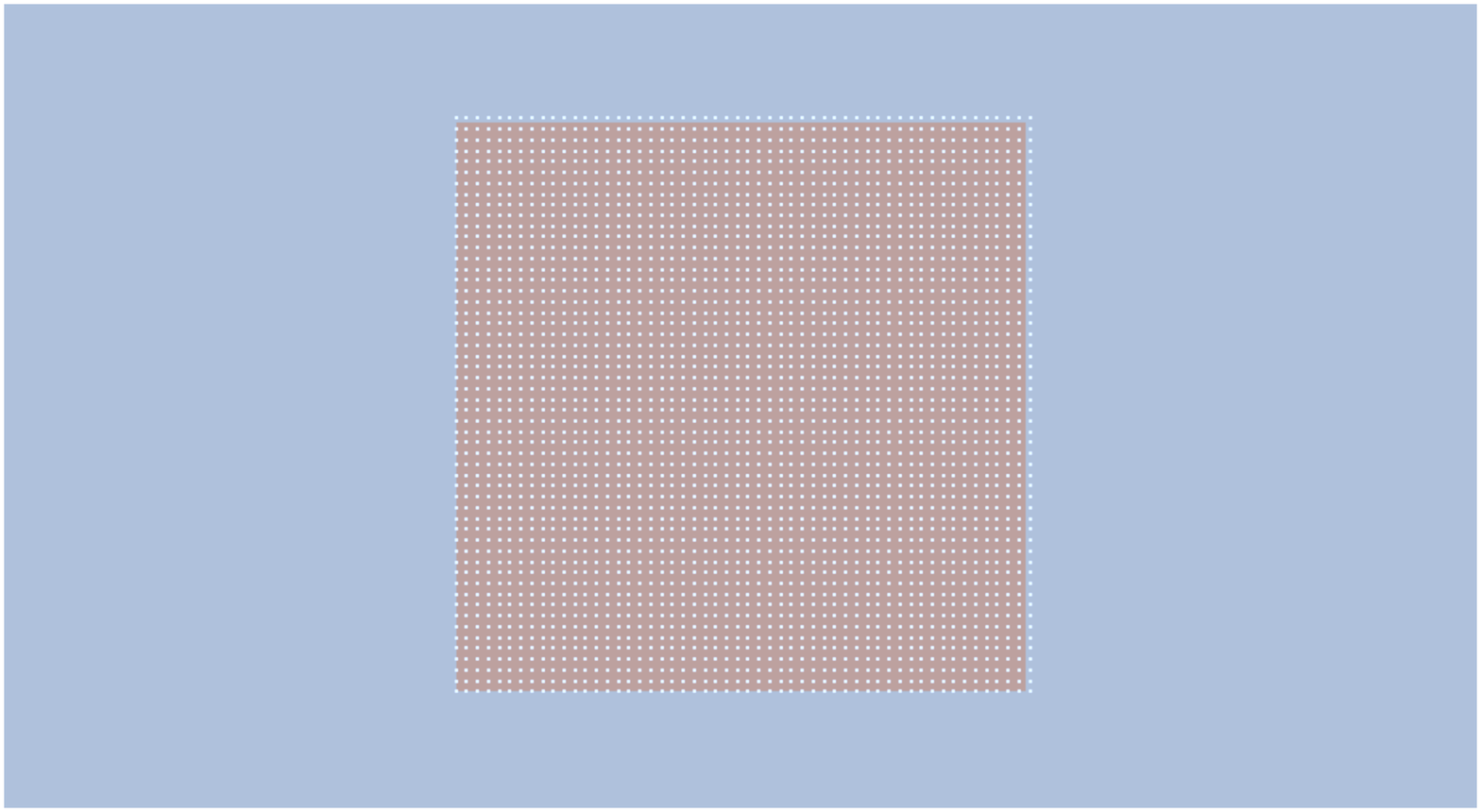}
\includegraphics[bb=135 280 451 582,clip,angle=90,width=0.39\columnwidth]{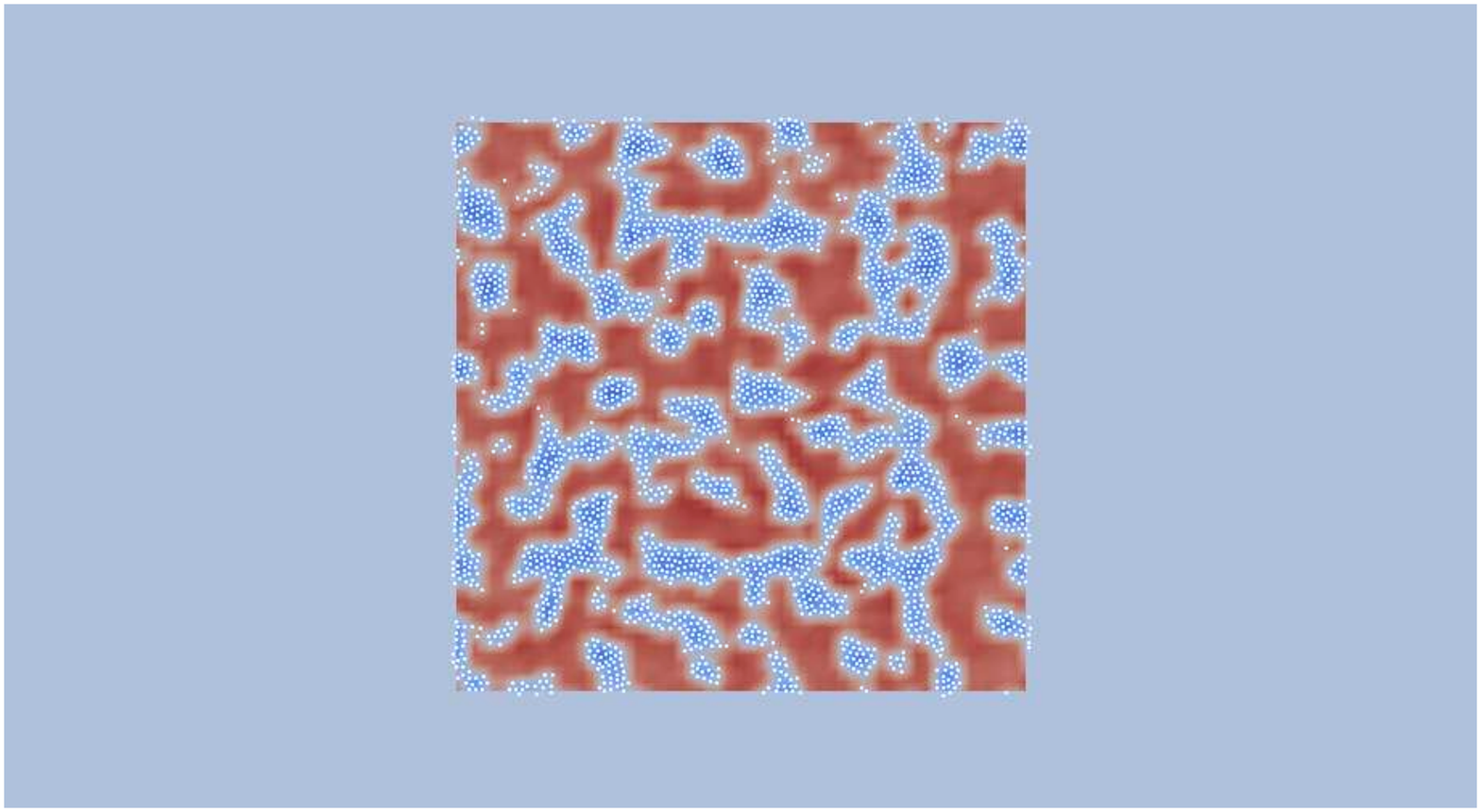}\par
\includegraphics[bb=135 280 451 582,clip,angle=90,width=0.39\columnwidth]{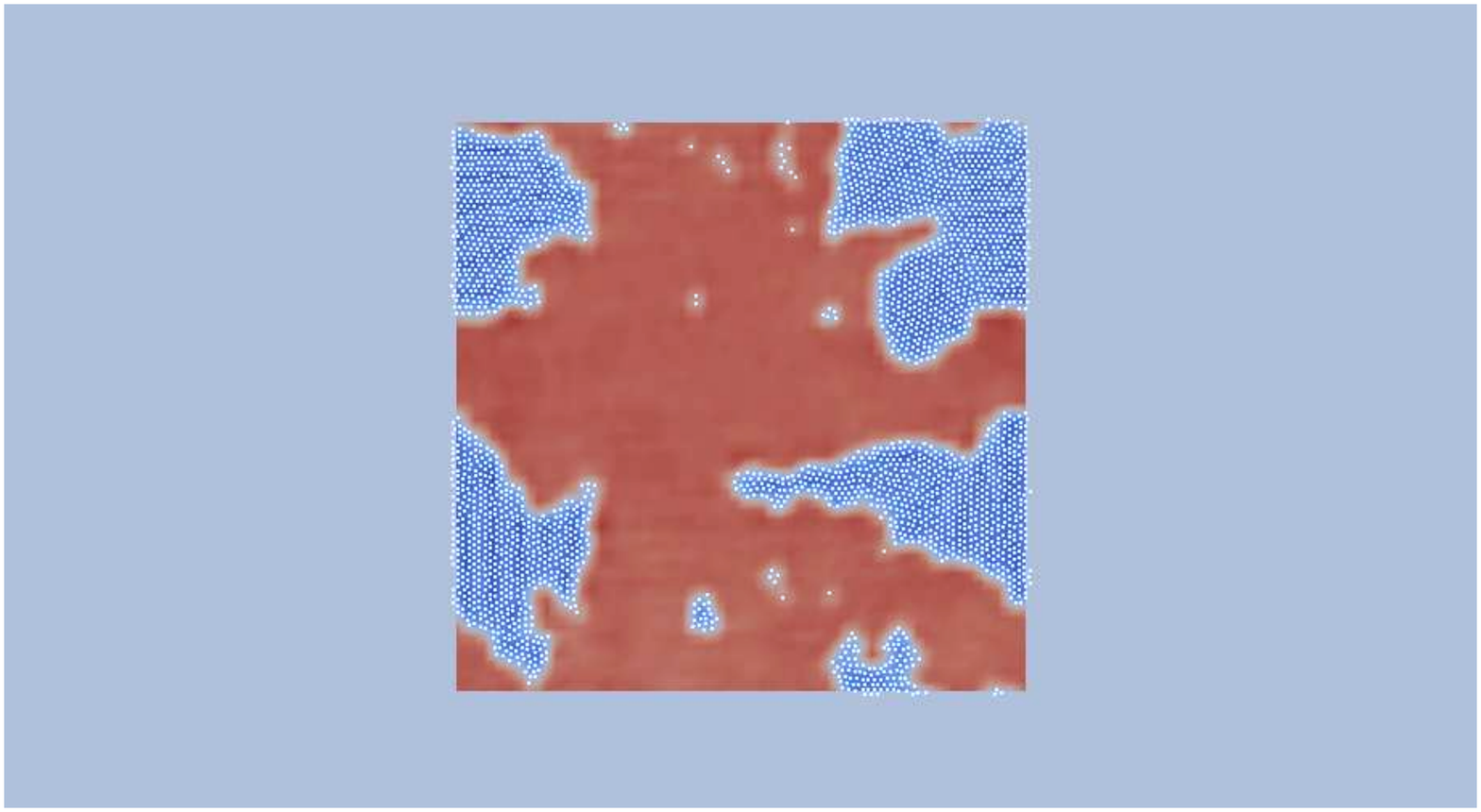}
\includegraphics[bb=135 280 451 582,clip,angle=90,width=0.39\columnwidth]{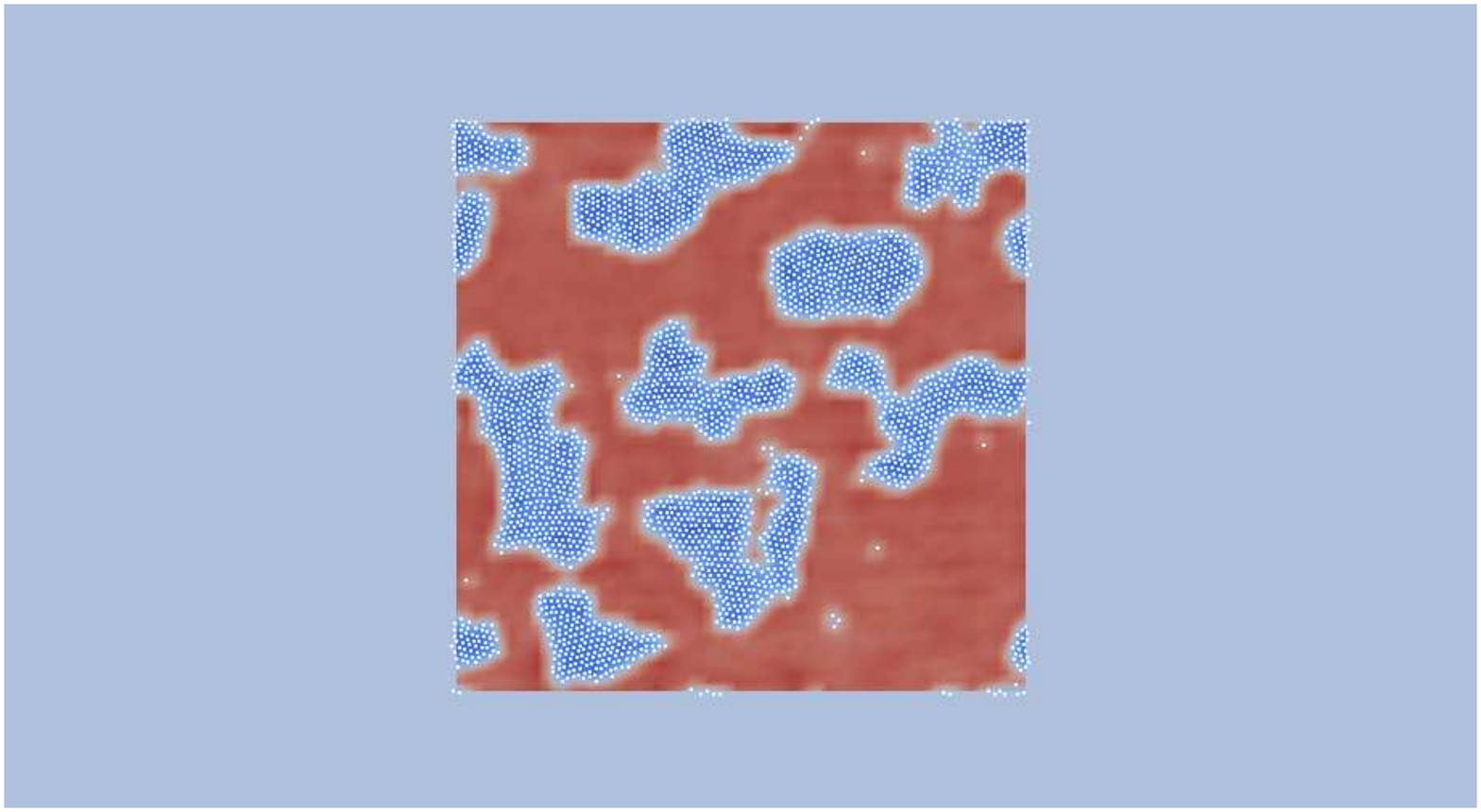}\par
\end{centering}
\caption{Four snapshots from a simulation of only one fluid component and a set of particles (time increasing from the top-left panel, clockwise). The fluid-particle system is able to separate, showing the particle-fluid symmetry.\label{multicomponent}}
\end{figure}

In Fig.~\ref{multicomponent} the evolution of
an initially homogeneous single component fluid in presence of an
ensemble of particle is shown. The fluid is simulated on a
$64\times64\times1$ grid with lattice spacing $a=1$, at a reduced
temperature $k_BT=1$. The only interaction terms are an excluded
volume interaction of the Weeks--Chandler--Anderson (WCA) type
between the particles, \begin{equation} U^{WCA}_{ij}(r_{ij})=
\begin{cases} 4\epsilon\left[
\left(\sigma/r_{ij}\right)^{12}-\left(\sigma/r_{ij}\right)^{6}
\right] & r_{ij}<2^{1/6}\sigma\\ 0 & r_{ij} \ge 2^{1/6}\sigma
\end{cases}, \end{equation} with parameters $\epsilon=0.1 k_BT$ and
$\sigma=0.7 a$, and the solvation free energies, equations
(\ref{eq:solvation_kappa}) and (\ref{eq:solvation_lambda}) with
coupling constants $\kappa=15 k_BT a^3$ and $\lambda=-110 k_BT a^3$. The
particles start grouping into small droplets, that eventually
coalesce into larger one, and a dynamic equilibrium between droplets
of different size, with continuous coalescence and breakup processes,
is attained.  This effect can not be achieved by means of the first
solvation term only, equation (\ref{eq:solvation_kappa}), as in
presence of an homogeneous fluid the solvation force
${\bm{F}}^{\mathrm{ps}}$ would be negligible. In this way, the
interactions can be tuned so that an ensemble of particles will
behave much like another fluid phase.  
The relatively low values
of Lennard-Jones interaction energy and particle radius, as well as the high values for both $\kappa$ and $\lambda$ values proved to
be necessary to achieve the demixing.  
Notice that while the particles are completely separating, the same
is not true for the fluid, that keeps a non-zero density also in
the particles-rich regions.
For the same purpose,
the choice of the solvation forces constants is not completely
independent from temperature, particle density and fluid density:
the absolute value of $\kappa$ needs to be large enough to prevent
particles from diffusing too much in the fluid due to thermal
fluctuations, therefore inducing demixing, and at the same time
$\lambda$  (which is responsible for fluid depletion in particle-rich
regions) should be kept small enough not to generate negative fluid
densities.  In other words, in this case  parameters $\kappa_{\zeta}$
and $\lambda_{\zeta}$ in (\ref{eq:solvation_kappa}) and
(\ref{eq:solvation_lambda}) play a purely phenomenological role and
one can use them to gauge the importance of the feedback of the
particles on the fluid evolution and vice-versa.  This goes together
with the idea of finding a proper renormalization of the average
feedback, such as to be able to describe realistic particles
concentration with only a reasonable number of them.
When using the solvation free energy
interaction~Eq.(\ref{eq:solvation_lambda}), care has to be taken
when using large values of $\lambda$, as they can induce strong
depletion in the nodes next to the particle, possibly ending up
with negative fluid densities and consequent failure of the Shan
Chen algorithm.

\section{Examples} In this section we present a series of examples
demonstrating how this approach can be used to study a wide class
of challenging problems. The Shan-Chen LB and the fluid-particle
coupling as described in the previous section has been implemented
in the ESPResSo software package
\cite{limbach2006espresso,arnold2013espresso}.  Thanks to the
flexibility, broad supply of interparticle potentials and methods
for the computation of electrostatic and magnetic properties with
different boundary
conditions\cite{arnold02a,arnold02c,arnold05b,tyagi07a,cerda08,tyagi08a,tyagi10a}
offered by ESPResSo, a broad range of systems can be modeled in an
effective way.

In all examples we will consider a binary mixture of two fluids
(say, $A$ and $B$).  We will discuss some issues associated with
the modeling of the contact angle at the interface between the two
fluids,  the interfacial deformations when colloidal particles are
crossing the dividing surface between two components, and the
surfactant effect of added amphiphilic molecules. We finally discuss
complex solutes simulated with flexible polyelectrolytes and explicit
counterions, and a case of ferrofluid emulsion.

\subsection{Modelling the Contact Angle} The effect of the solvation
force, equation (\ref{eq:solvation_kappa}) is to drive a particle
along the direction of the density gradient of the fluid component.
If the coupling constants $\kappa_\zeta$ for the two fluids have
opposite sign, the particle will simply move towards the maximum
(or the minimum, depending on the sign of the interaction) of one
of the two components. If the particle is instead repelled by both
components (i.e., both constants are positive), it will be driven
to the interface, and its equilibrium position on the difference
between the two forces.  

\begin{figure}
\begin{centering}
\includegraphics[width=0.9\columnwidth]{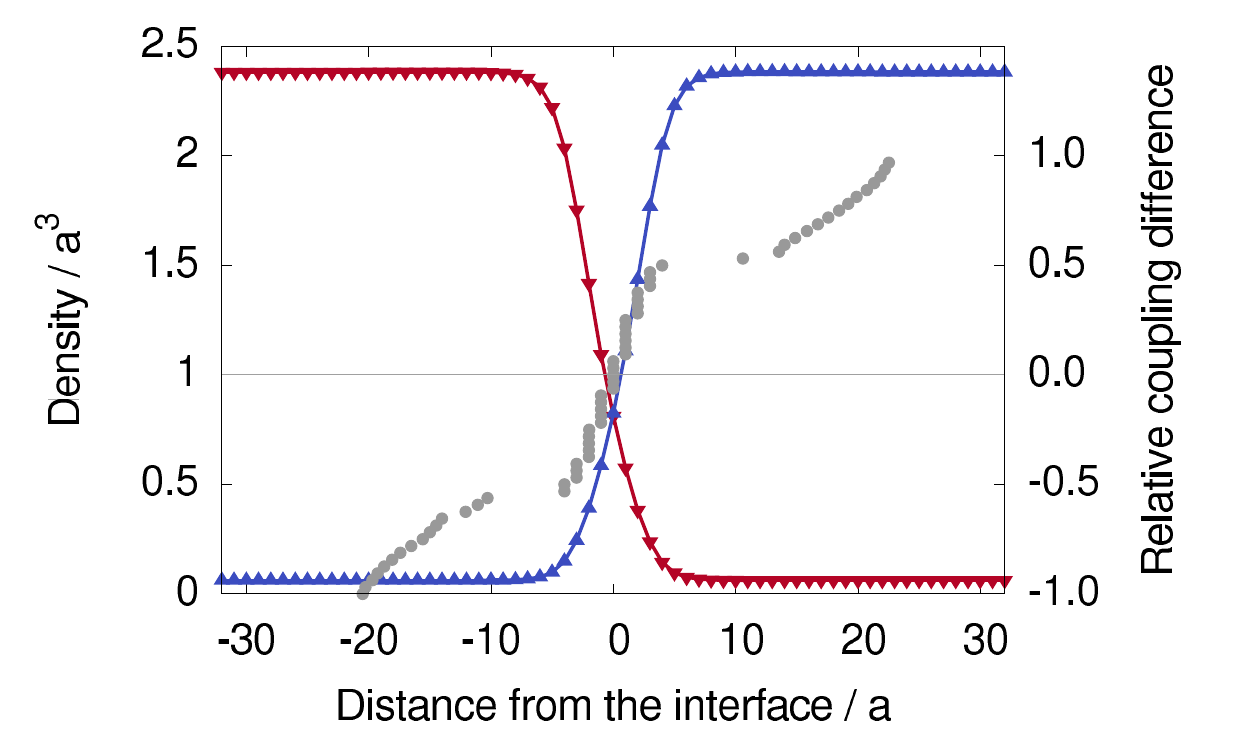}
\includegraphics[width=0.9\columnwidth]{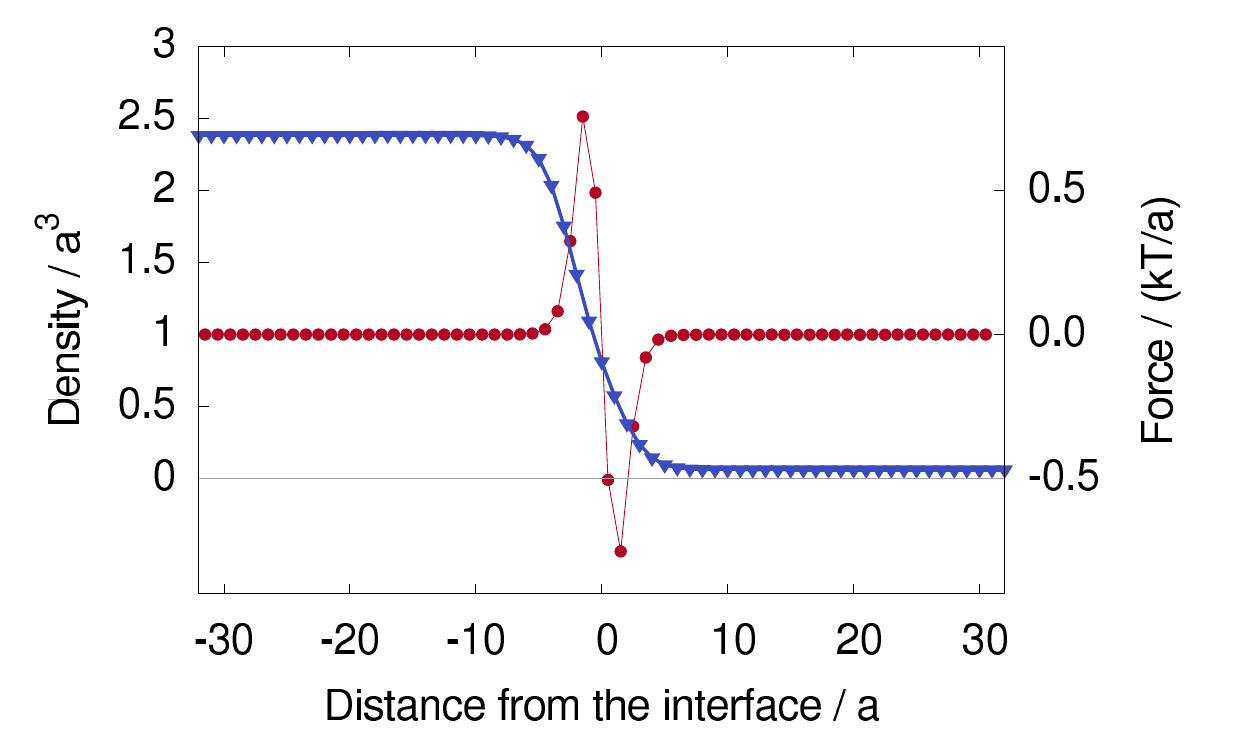} 
\end{centering}
\caption{Upper
panel: density profiles of the $A$ (lower triangles) and $B$ (upper
triangles) components as a function of the distance from the
interface. Circles denote the equilibrium distance for a given
relative coupling difference $2 (\kappa_A-\kappa_B)/(\kappa_A+\kappa_B)$
(right vertical axis). Lower panel: density profile of the $A$
component (lower triangles) and force exerted on a particle that
is kept fixed at a given distance from the interface
(circles).\label{fig:contact}} \end{figure}

In general, the ability of a solid particle $S$ to adsorb to a given
interface between fluid $A$ and fluid $B$ is determined by a balance
of surface forces. When the two solid-fluid tensions ($\gamma_{SA}$,
$\gamma_{SB}$) are different, the lowest free energy state has the
particle on the interface so long as the contact angle $\theta$
satisfies $\gamma_{AB} \cos \theta=\gamma_{SA}-\gamma_{SB}$
($\gamma_{AB}$ is the surface tension of the liquid-liquid interface),
with $0 \le \theta \le \pi$ (partial to complete wetting). In the
absence of body forces on the particle, the interface remains
perfectly flat while the particle is displaced so that it intersects
the interface at the angle $\theta$. When dealing with point
particles, however, a thermodynamic equivalent to the particle
radius has to be introduced, in order to define an effective contact
angle. The force $F_D$ required to detach a spherical particle of
radius $R_{\mathrm{eff}}$ from the interface is \be F_{D}=-\gamma_{AB}
\pi R_{\mathrm{eff}} (1 \pm |\cos \theta|), \ee where the $+(-)$
sign applies to a particle being pulled out of (into) its preferred
solvent \cite{Pieranski80,BinksHorozov06}. By choosing the coupling
constants $\kappa_A = \kappa_B > 0$,  a particle adsorbs exactly
in the middle of the diffuse interface (see Fig.~\ref{fig:contact}),
defining a reference state with $\theta=\pi/2$. The maximum of the
force acting on the particle as it crosses the interface
(Fig.~\ref{fig:contact}, lower panel) corresponds to $F_{D}$, and
its value allows to estimate the effective radius of the particle,
$R_{\mathrm{eff}} = F_D / (\gamma_{AB} \pi)$ and, consequently, the
cosine of the contact angle $\cos \theta = d/R_{\mathrm{eff}}$ as
a function of the equilibrium distance $d$ of the particle from the
surface. For the case reported in Figures \ref{fig:contact},
$\gamma_{AB}=0.132$ and $R_{\mathrm{eff}}\simeq1.8$.

\subsection{Colloid crossing an interface.} 
\begin{figure}
\begin{center} (a)\includegraphics[bb=185 0 400
842,clip,angle=270,width=0.62\columnwidth]{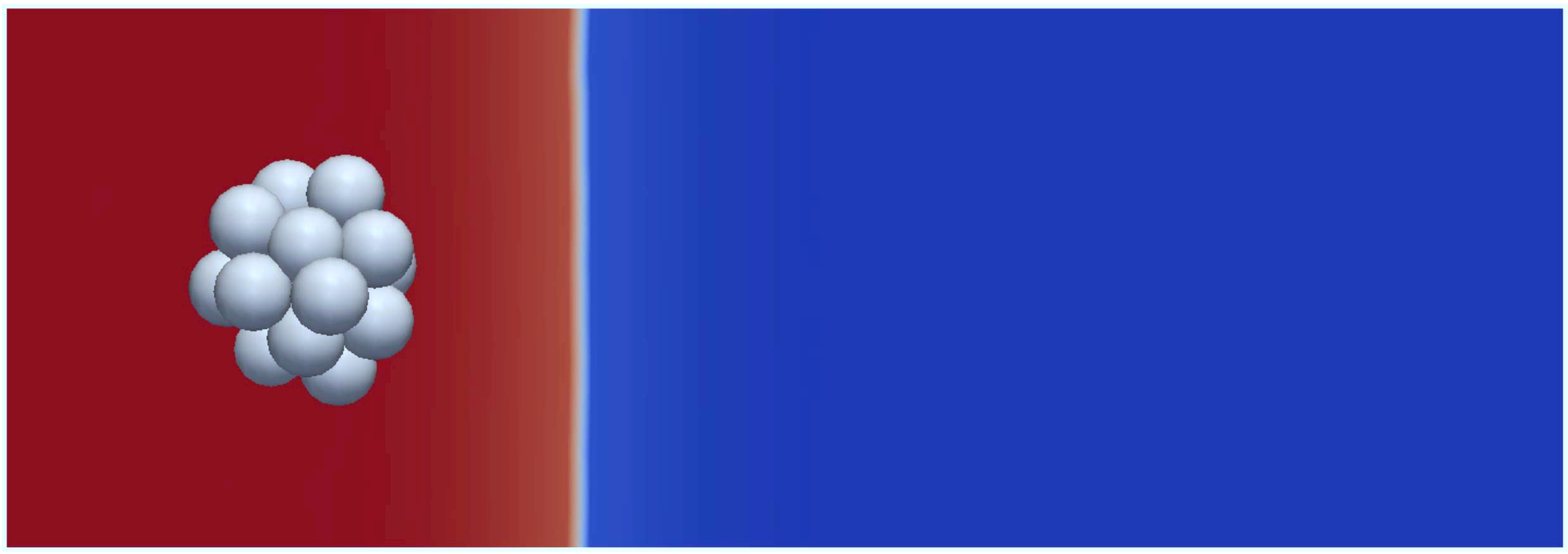}
(b)\includegraphics[bb=185 0 400
842,clip,angle=270,width=0.62\columnwidth]{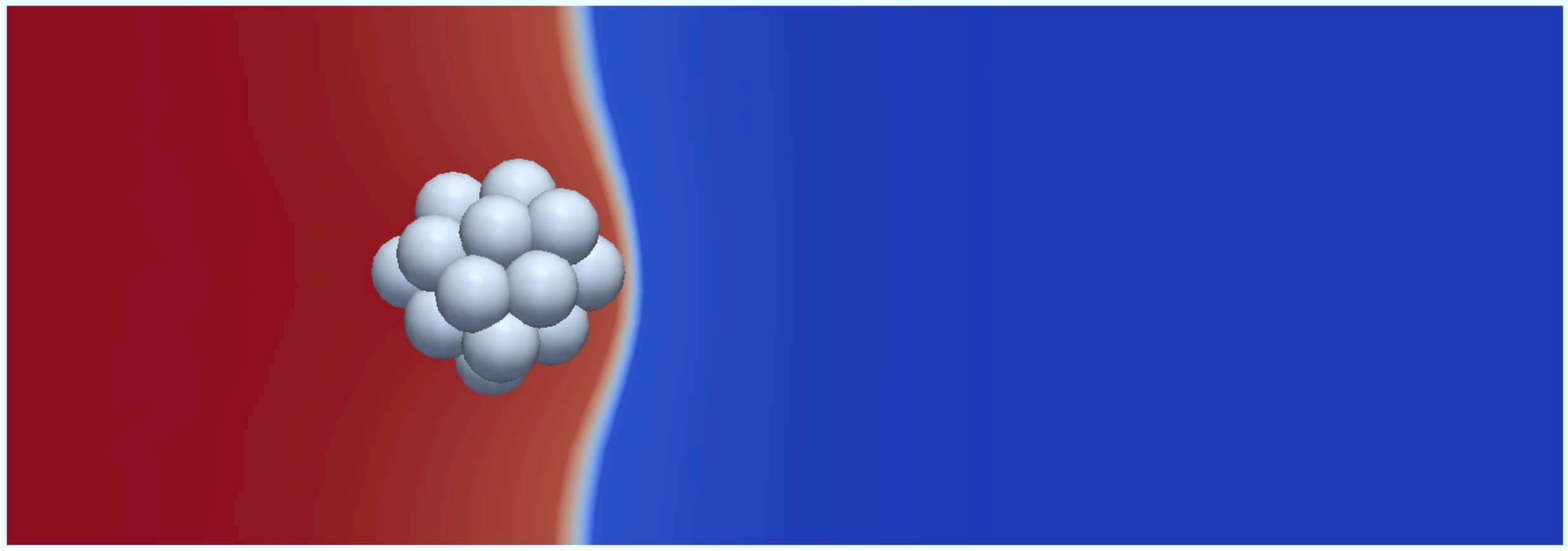}
(c)\includegraphics[bb=185 0 400
842,clip,angle=270,width=0.62\columnwidth]{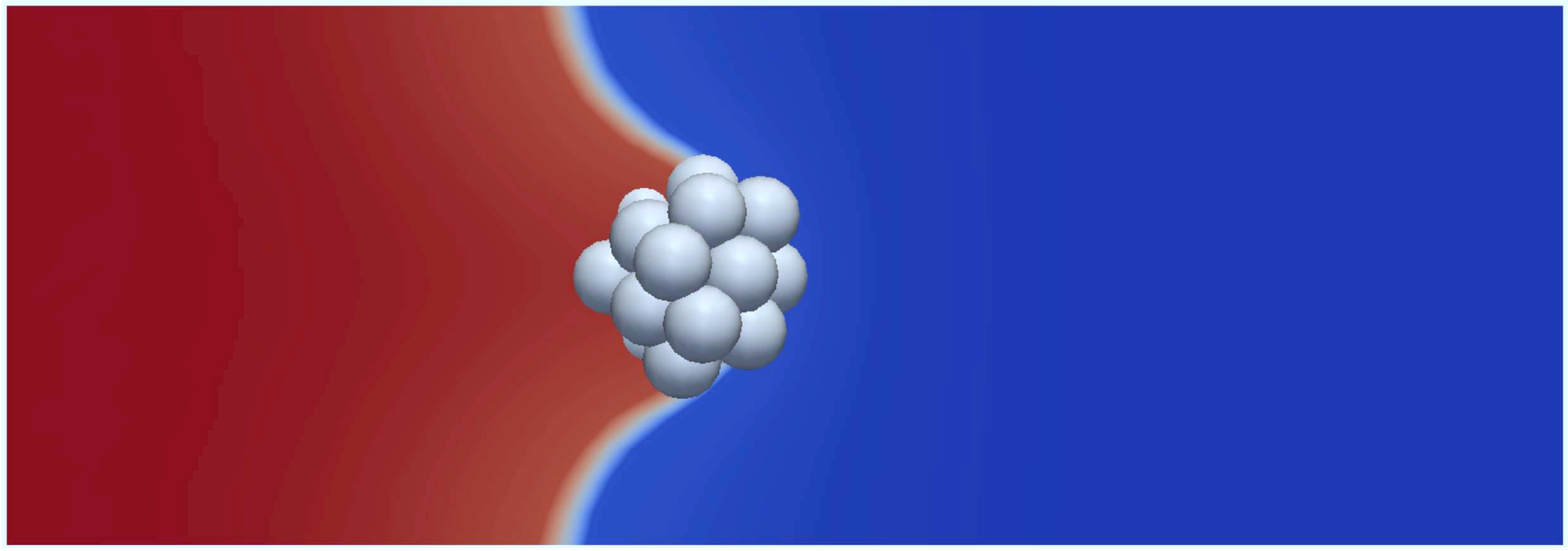}
(d)\includegraphics[bb=185 0 400
842,clip,angle=270,width=0.62\columnwidth]{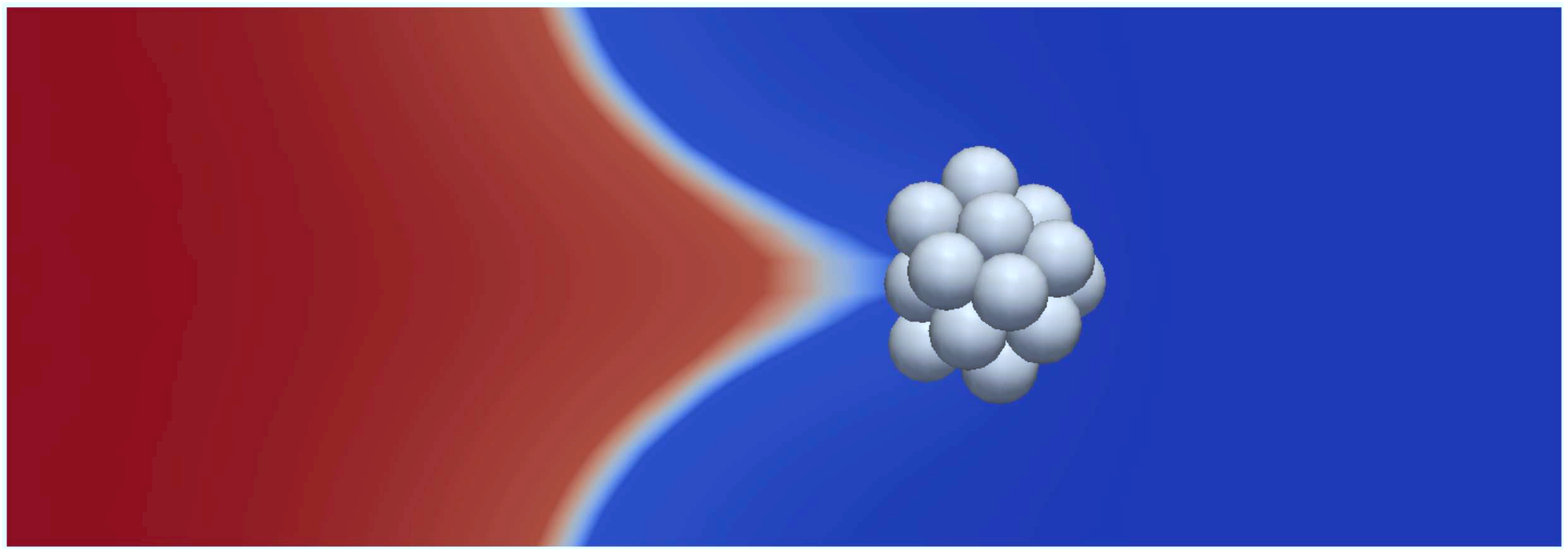}
(e)\includegraphics[bb=185 0 400
842,clip,angle=270,width=0.62\columnwidth]{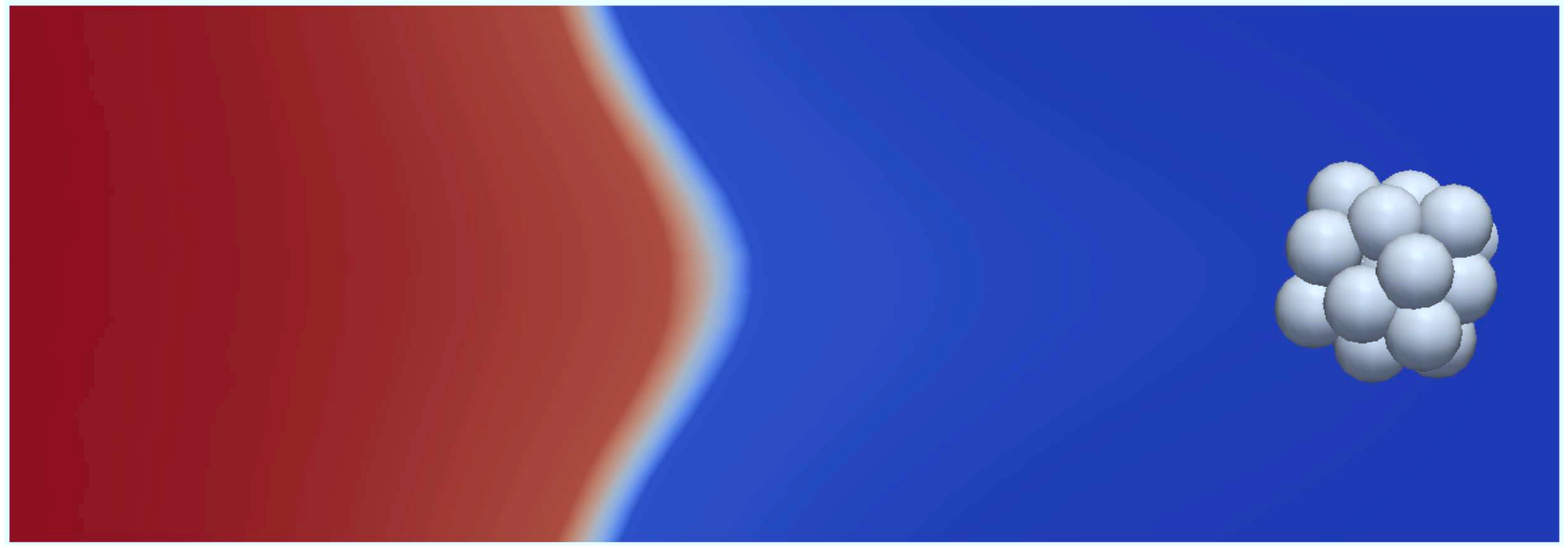} 
\end{center}
\caption{A raspberry model colloid is pushed through the interface
of two fluids by applying a constant force on it.\label{raspberry}}
\end{figure}

To show the effect of
the coupling term in equation (\ref{eq:solvation_lambda}), we present
the results of the simulation of a raspberry \cite{lobaskin04} model
colloid pushed with constant force through the interface between
two fluid components (Fig.~\ref{raspberry}).  Both fluid components
have an average density of 118.0 (to mimic a high surface tension)
and Shan-Chen off-diagonal coupling terms $g_{12}=0.023$, which
produce a macroscopic demixing of the two fluids.  The particle-fluid
coupling constant are $\kappa_A=\kappa_B=0$ and
$\lambda_A=-\lambda_B=20 k_BT a^3$. The choice of the $\lambda_\zeta$
parameters makes one of the fluid component accumulate around the
particles, while the other one is pushed away.  

The effect of the $\lambda_\zeta$ coupling constant is clearly seen
in the snapshot (b) of Fig.~\ref{raspberry}, where the interface
starts being deformed by the colloid as soon as the first beads
reach the dividing surface (white line). Then, the deformation of
the surface keeps extending until a contact angle of about 45 deg.~is
reached (c).  At this stage the elastic energy of the interface
arising from its surface tension is roughly equivalent to the
effective solvation energy of the colloid. After further displacement
(d), the solvation force is not able to sustain the surface tension
anymore, and the colloid detaches from the interface, which eventually
(e) relaxes back towards its flat, equilibrium shape.

\subsection{Modeling amphiphilic molecules as surfactants.}
\begin{figure} \begin{center} 
\includegraphics[bb=120 130 480 680,clip,width=0.30\columnwidth]{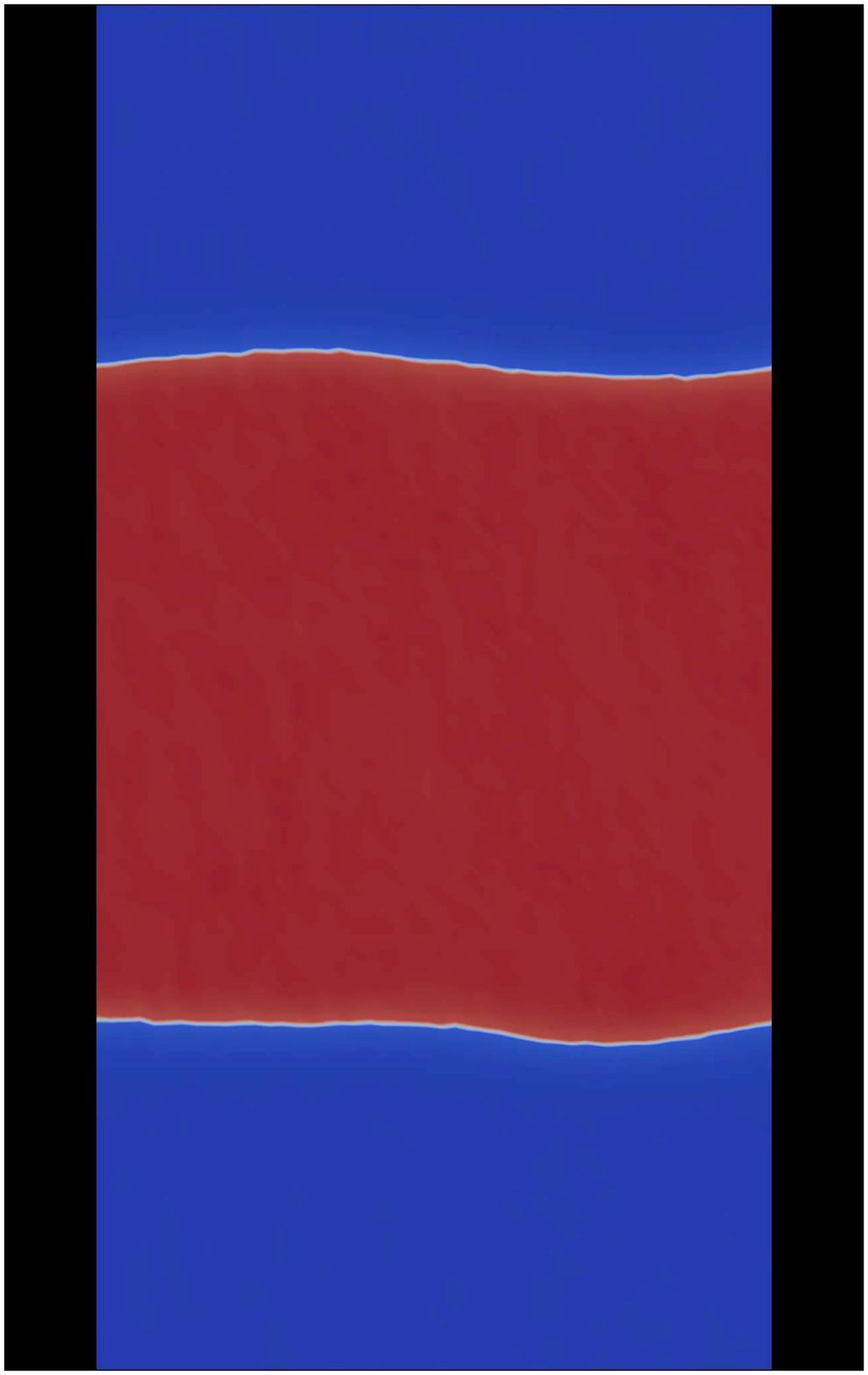} 
\includegraphics[bb=120 50  480 600,clip,width=0.30\columnwidth]{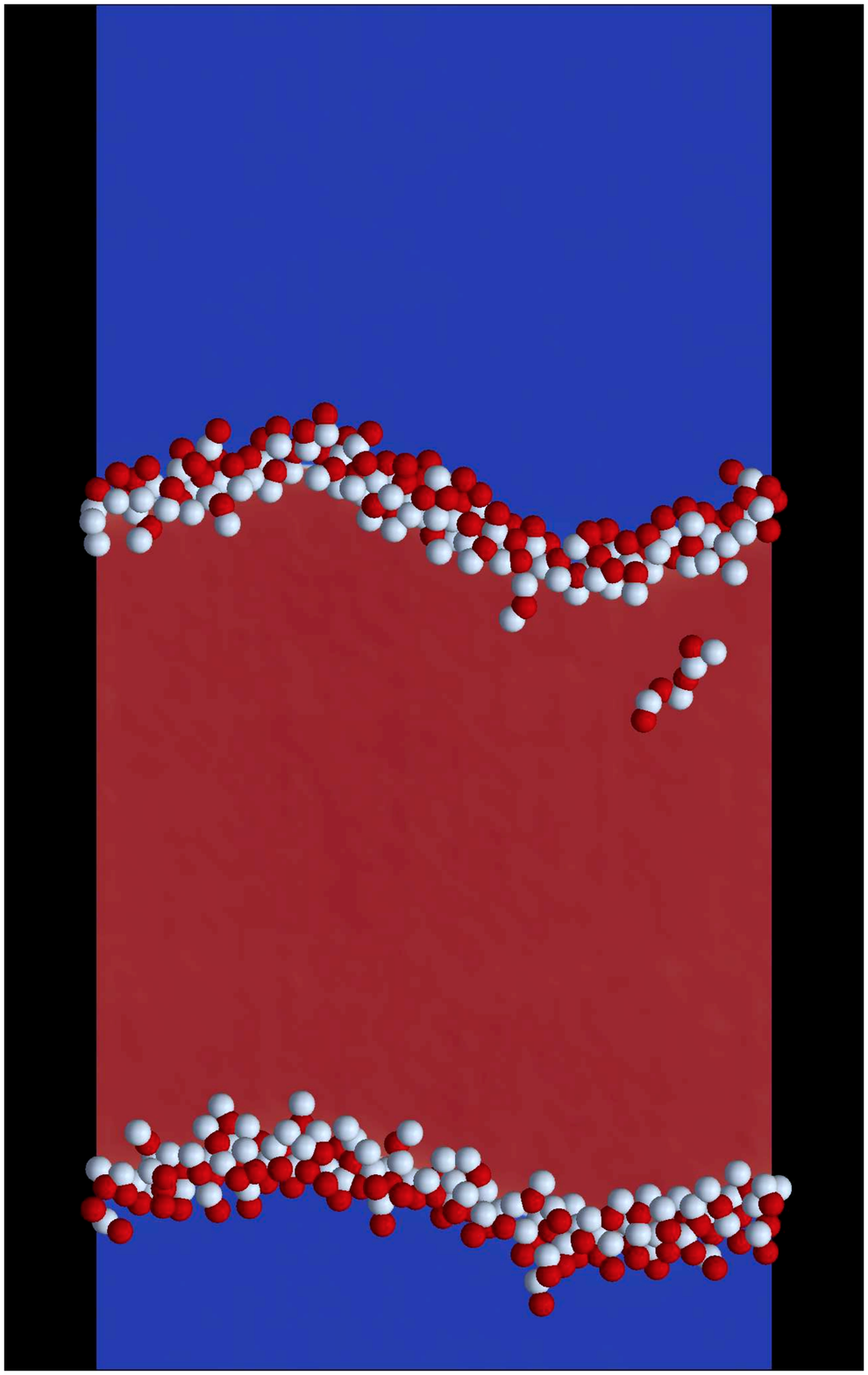}
\includegraphics[bb=0 0 400 250,clip,width=0.90\columnwidth]{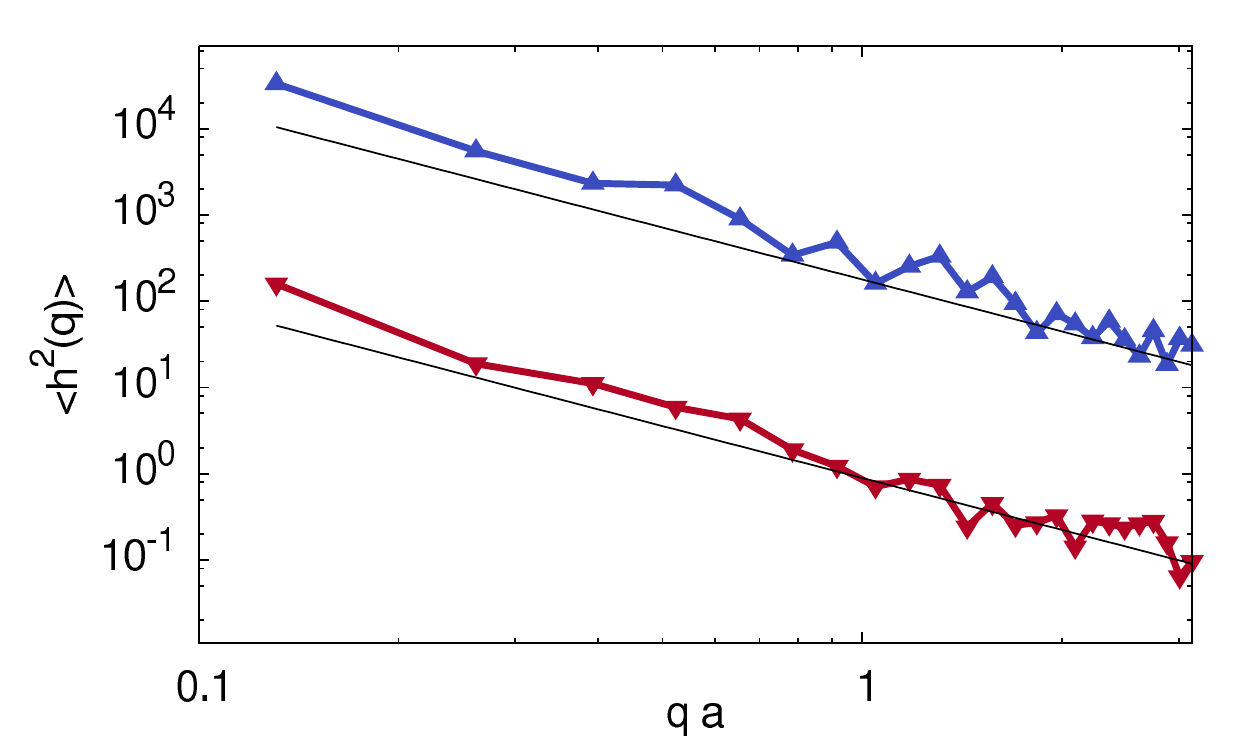}
\end{center} \caption{Upper panel: simulation snapshots 
showing capillary waves induced by thermal fluctuations.
In the pure bicomponent fluid (left) the fluctuations are smaller
than in presence of surfactant molecules (right). Head and tail beads 
are depicted as red and white spheres, respectively. Lower panel: the 
surface fluctuations spectra as a function of inverse distance
(the dataset with smaller fluctuations refers to the pure liquid).
\label{surftens}}
\end{figure}
The example of the raspberry colloid crossing the interface has shown that the interaction Eq. (\ref{eq:solvation_lambda}) can be used to induce deformations in the interface. This suggests that the particle-fluid coupling could be used to model the surfactant action of amphiphilic molecules. We simulated model amphiphilic molecules in a ($A,B$) bicomponent fluid on a $96\times48\times2$ grid with spacing $a=1$, at reduced temperature $k_BT=1$. Coarse-grained surfactants composed of one ``head'' and one ``tail'' beads connected
by a harmonic spring (with spring constant $k_{\mathrm{har}}=20
k_BT/a^2$ and equilibrium distance $d=0.8 a$) are modeled using
$A$-philic and $B$-phobic interactions ($\kappa_A=-\kappa_B=-0.5 k_BT a^3$)
for the tail beads and vice versa ($\kappa_B=-\kappa_A=-0.5 k_BT a^3$) for
the head beads. Additionally, the head beads are acting on the $A$
fluid component using the force Eq.~(\ref{eq:solvation_lambda})
with $\lambda_A= -k_BT a^3$. The excluded volume of the beads is modeled
using WCA interaction with parameters $\epsilon=4 k_BT$ and $\sigma=1.2
a$, between all pairs within the cutoff radius $2^{1/6}\sigma$.

The surfactant molecules are initially placed randomly in the
simulation box, and they quickly move to the interface, where they influence the underlying fluid profile. The surfactant
action of these model amphiphilic is evident from the comparison
of two typical snapshots (see Fig.~\ref{surftens}, upper panel)
of the fluid in presence and absence of the molecules themselves,
but can be quantified by looking at the spectrum of the interface
fluctuations. In the continuum limit, the local position $h(x)$ of
a single interface subject to thermal fluctuations has an average
spectrum $\left\langle h^2(q)\right\rangle$ that grows like the
inverse of $q^2$\cite{safran94}, \begin{equation} \left\langle h^2(q)\right\rangle=\frac{2
k_BT}{\gamma_{AB}}\frac{1}{q^2}, \label{eq:surftens} \end{equation} and
is also inversely proportional to the surface tension $\gamma_{AB}$. 

The spectrum of the fluctuations is shown in the lower panel of
Fig.~\ref{surftens}, where the constant value measured at high
$q$s (where the continuum approximation is not valid anymore) has
been subtracted. The solid, straight lines are the functions $k_BT
\gamma_{AB}^{-1} q^{-2}$ for two different values of $\gamma_{AB}$. The
straight line in correspondence with the data for the interface in
absence of surfactant is not obtained from a best fit procedure,
but by using the value $\gamma_{AB}=2.24$ obtained from an independent
set of simulations at $T=0$ of spherical droplets with different
radii $R$ by fitting Laplace's law (which relates the capillary
pressure jump $\Delta p$ across the interface to the surface tension)

\begin{equation} 
\Delta p =2\frac{\gamma_{AB}}{R}. 
\end{equation} 

The second straight line represent instead the best fit to the
theoretical expression, Eq. (\ref{eq:surftens}) for the
fluctuation spectrum in presence of surfactant, that leads to
$\gamma_{AB} \simeq 0.011$, namely a surface tension about 200 times lower
than in absence of surfactants.

\subsection{Polyelectrolytes, bicontinuous structure and electrostatic screening.}

As an example of a complex fluid-fluid interface, we simulated the
relaxation towards equilibrium of a mixture of polyelectrolytes and
their counterions in a two-components fluid. Both fluids start from
a homogeneous distribution with density $\rho_A=\rho_B=118/a^3$ with
Shan-Chen coupling terms $g_{AB}=g_{BA}=0.023$ on a $32\times32\times32$
grid with spacing $a=1$. The polymers (10 chains, each 64 monomers
long) are described using a bead-and-spring model with harmonic
constant $k_{\mathrm{har}}=0.5 k_BT/a^2$ and equilibrium distance
$0.8 a$. Every bead is interacting with the all other ones via a
WCA potential with $\sigma=a$ and $\epsilon=k_BT$. Each bead has
a unitary charge, in reduced units, $q_i=1$, and is neutralized by a counterion with
opposite charge, interacting with the same WCA potential as the
polymer beads. The electrostatic pair energy 
\begin{equation}
U^{ES}_{ij}(r_{ij})=\ell_B k_BT\frac{q_iq_j}{r_{ij}}, 
\end{equation}
with Bjerrum length $\ell_B=1$, is computed using the P3M
algorithm \cite{ewald21, hockney88, deserno98, deserno98a} taking
into account the presence of periodic copies in
all directions.  The solvation free energy parameters are
$\kappa_{A}=-\kappa{B}=-5\times10^{-4} k_BT a^3$, $\lambda_A=0.01 k_BT a^3$ and $\lambda_B=0$
for both polymer beads and counterions.

\begin{figure}[t]
\includegraphics[bb=142 180 500 542,clip,angle=90,width=0.32\columnwidth]{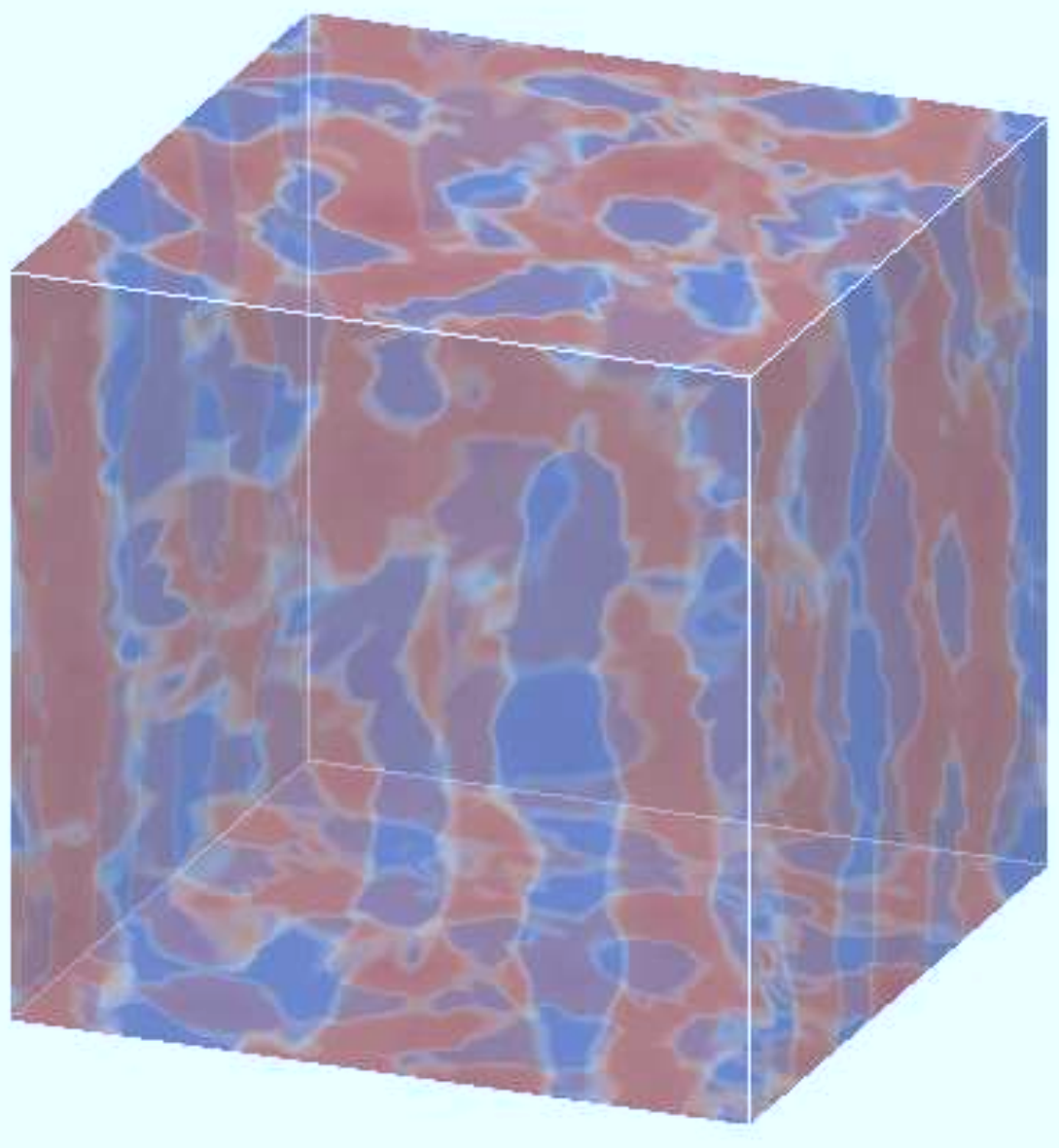}
\includegraphics[bb=142 180 500 542,clip,angle=90,width=0.32\columnwidth]{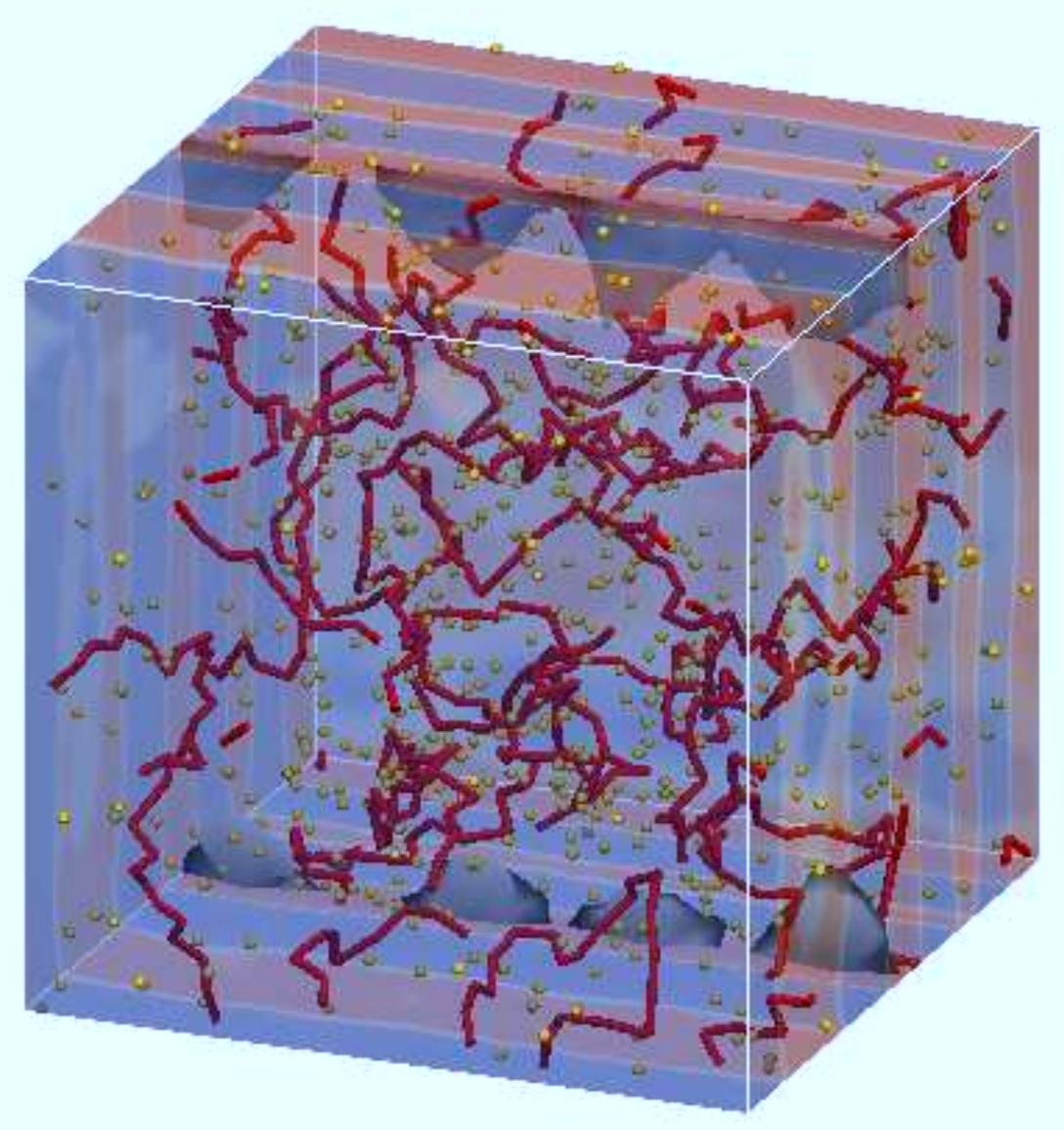}
\includegraphics[bb=142 180 500 542,clip,angle=90,width=0.32\columnwidth]{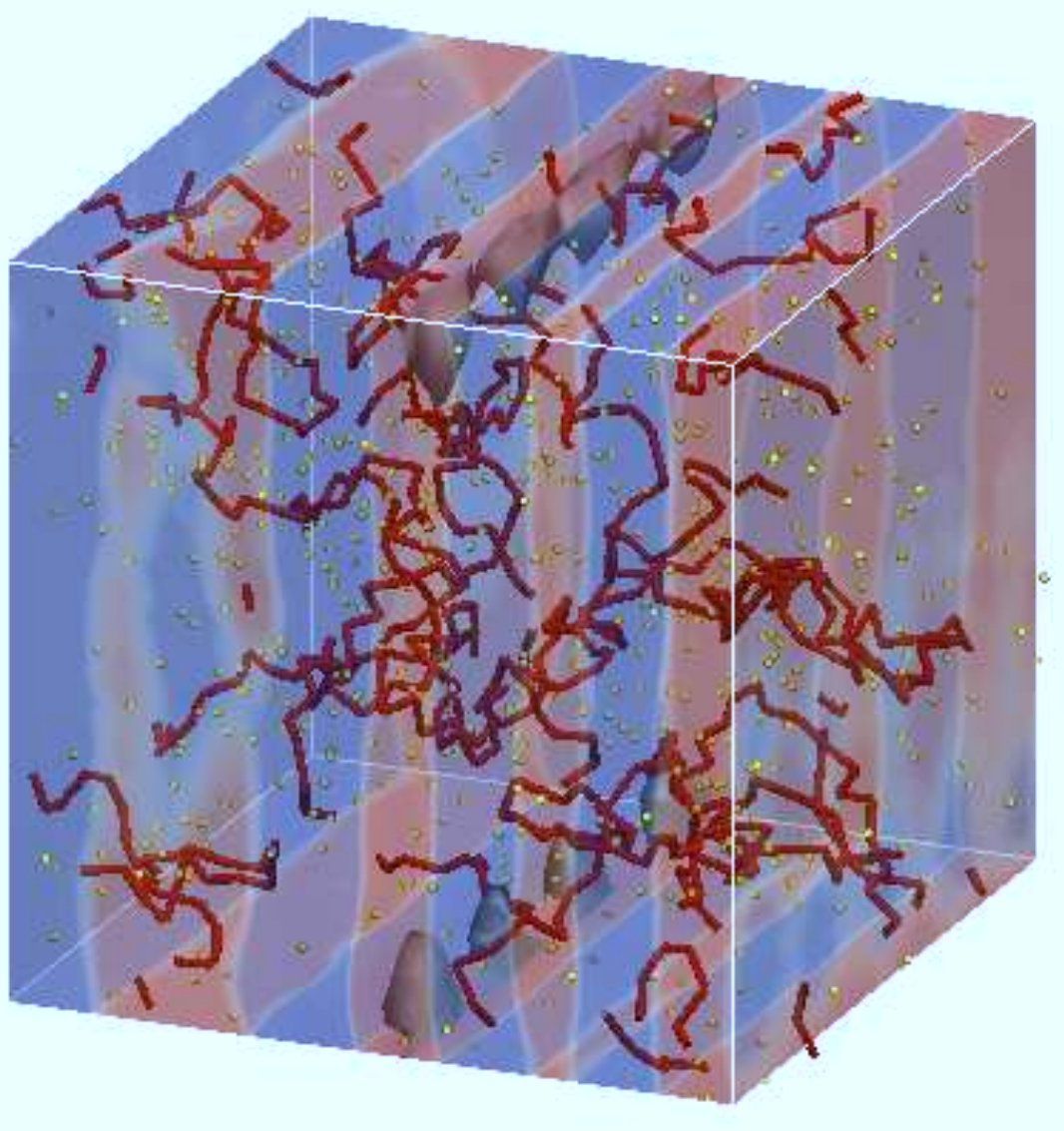}
\includegraphics[bb=142 180 500 542,clip,angle=90,width=0.32\columnwidth]{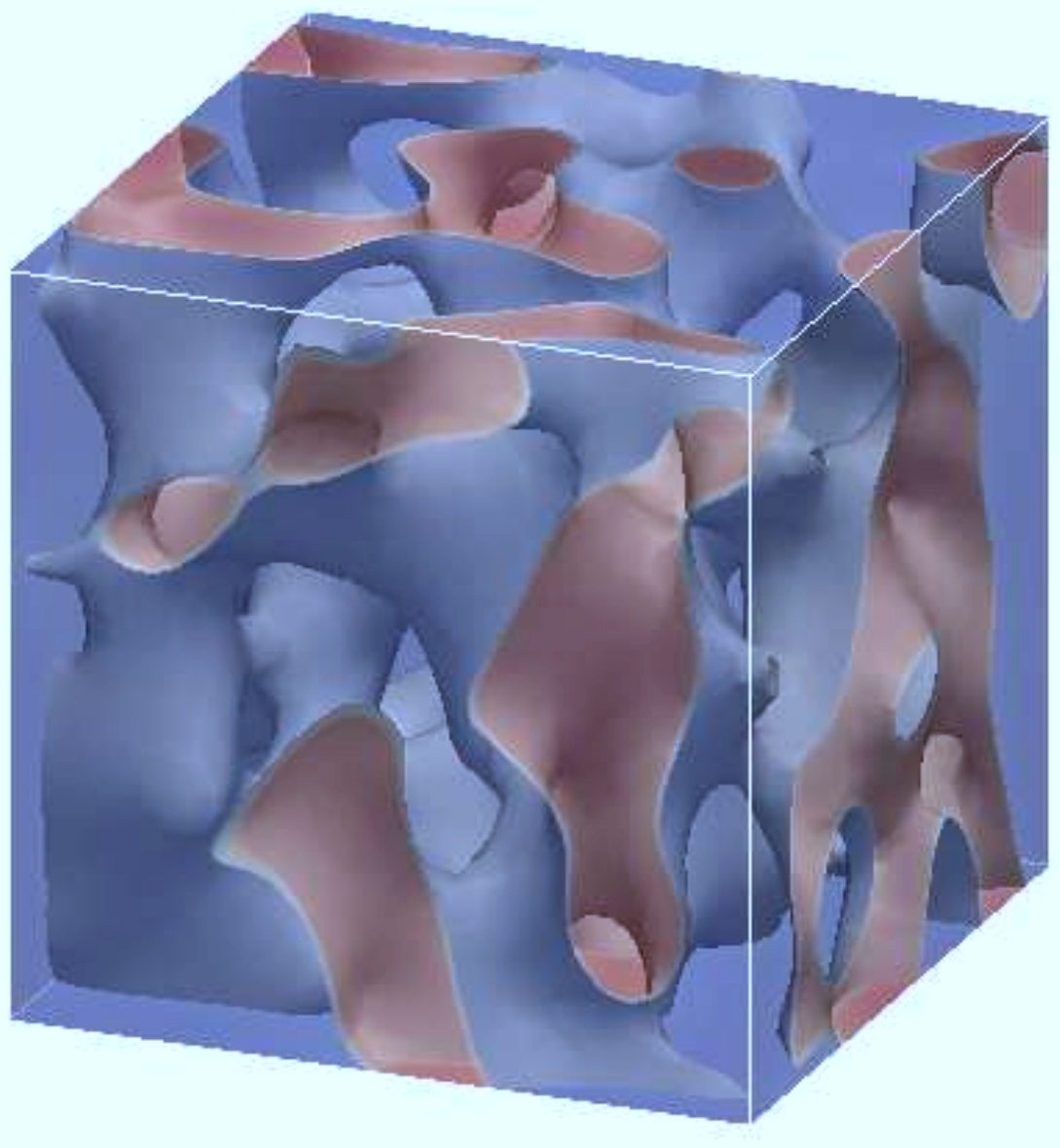}
\includegraphics[bb=142 180 500 542,clip,angle=90,width=0.32\columnwidth]{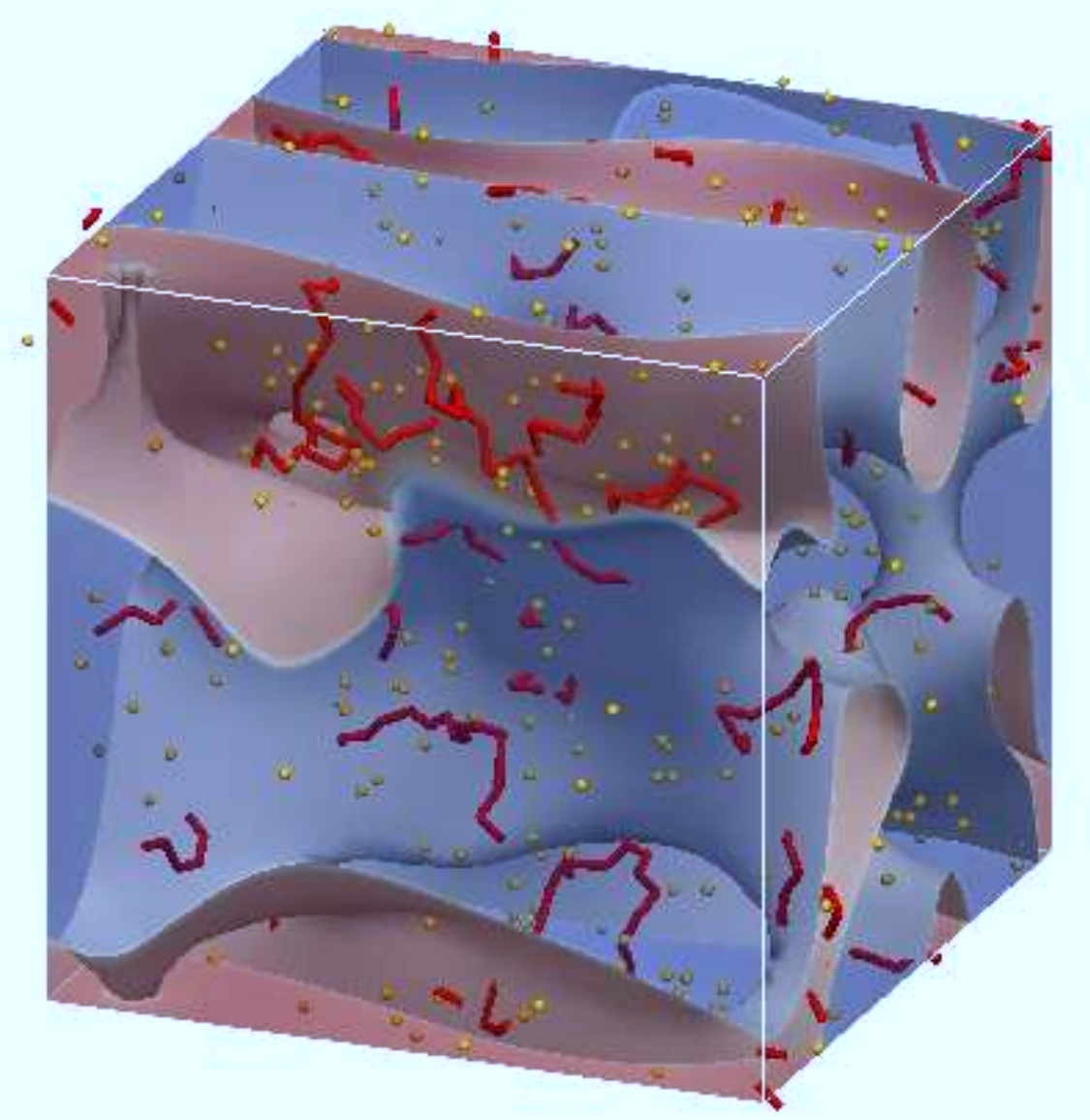}
\includegraphics[bb=142 180 500 542,clip,angle=90,width=0.32\columnwidth]{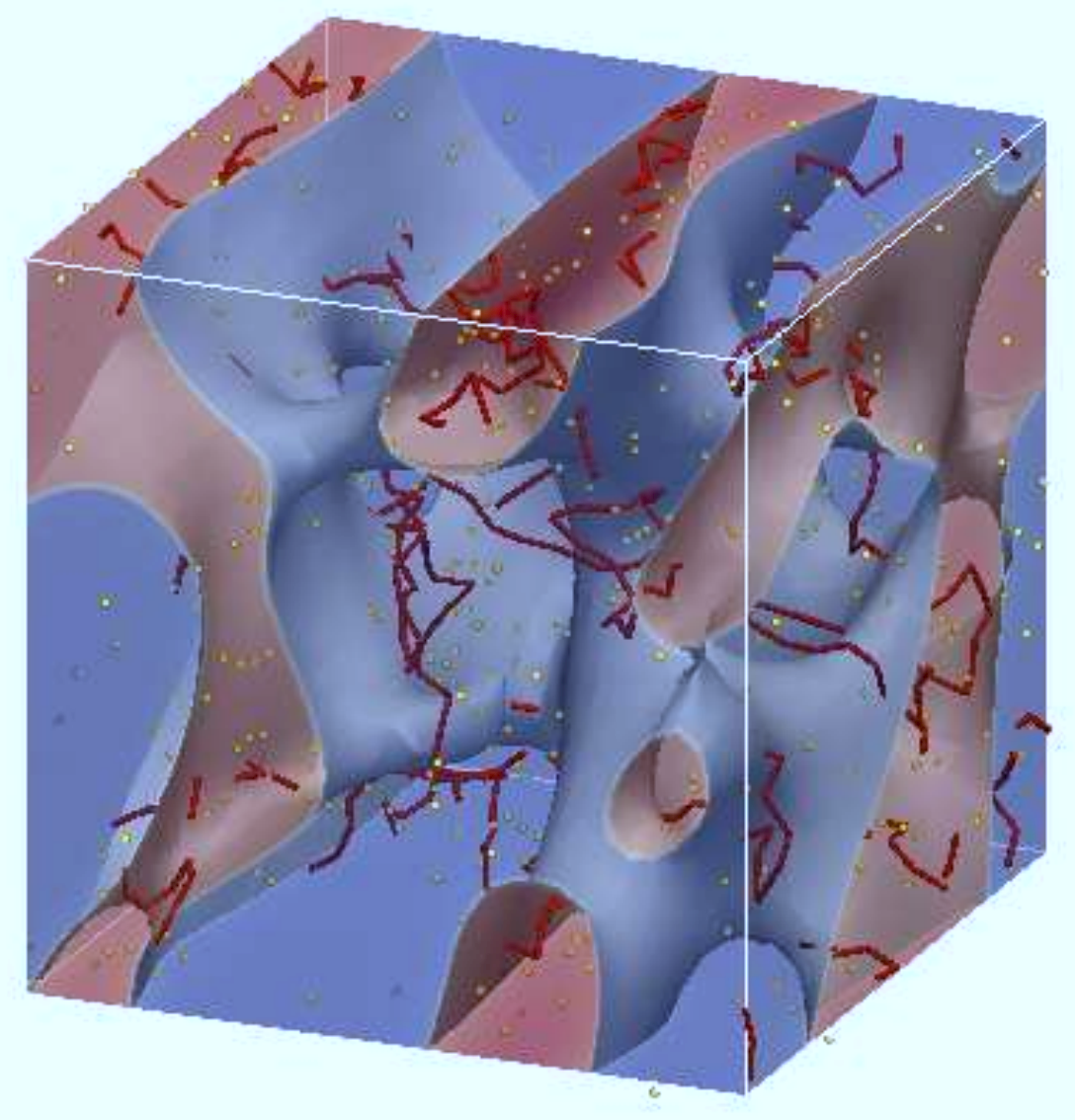}
\includegraphics[bb=142 180 500 542,clip,angle=90,width=0.32\columnwidth]{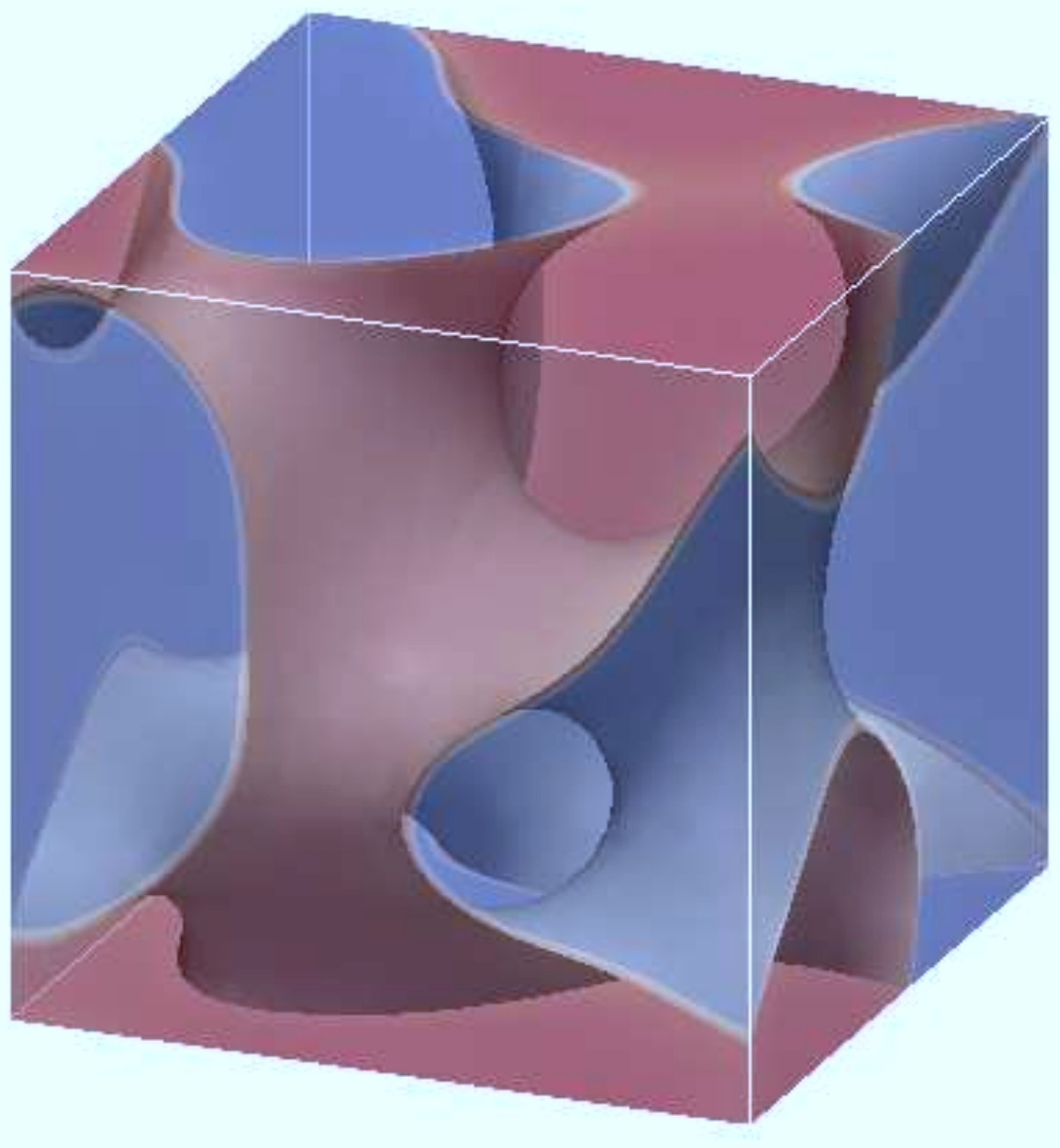}
\includegraphics[bb=142 180 500 542,clip,angle=90,width=0.32\columnwidth]{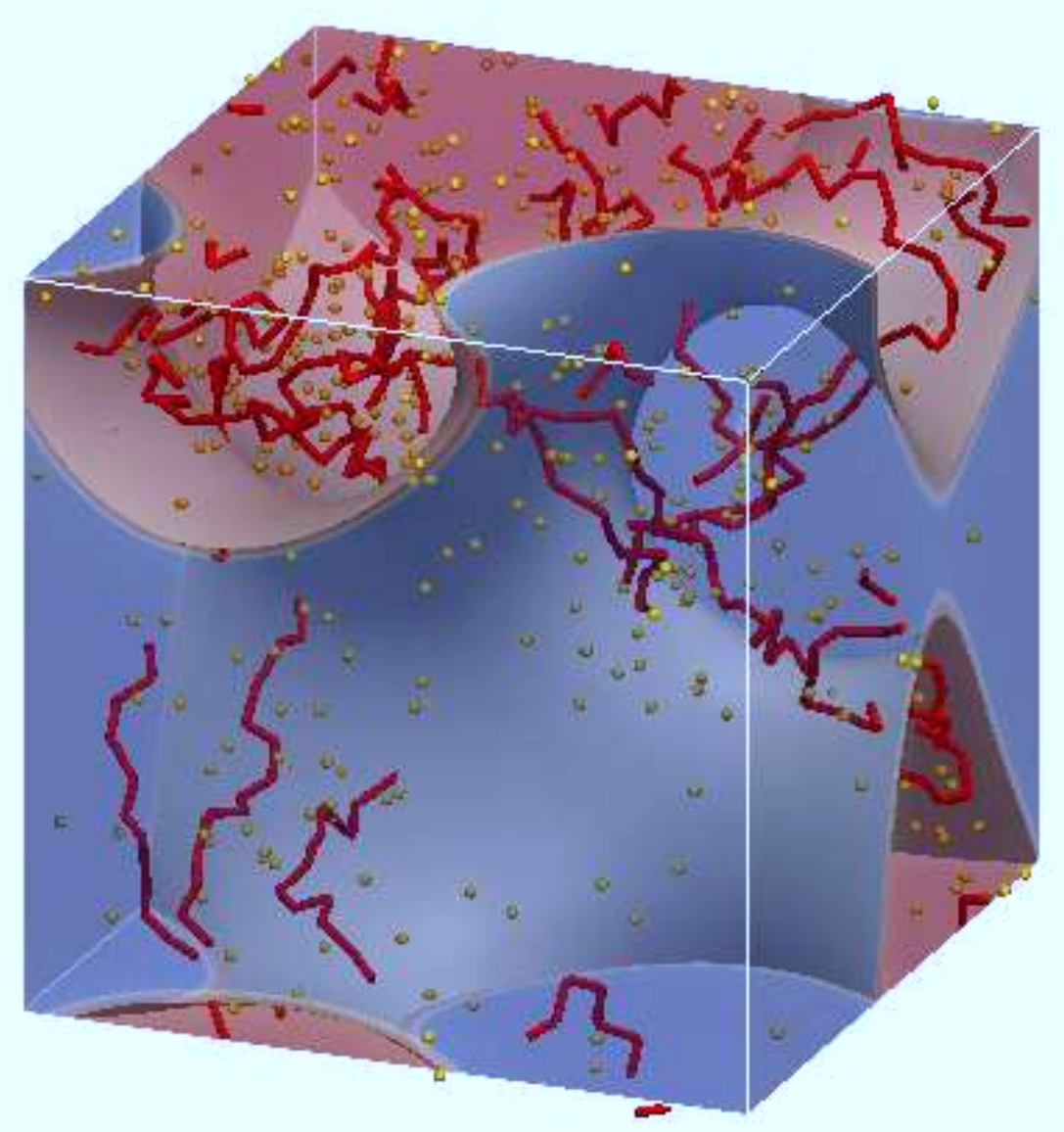}
\includegraphics[bb=142 180 500 542,clip,angle=90,width=0.32\columnwidth]{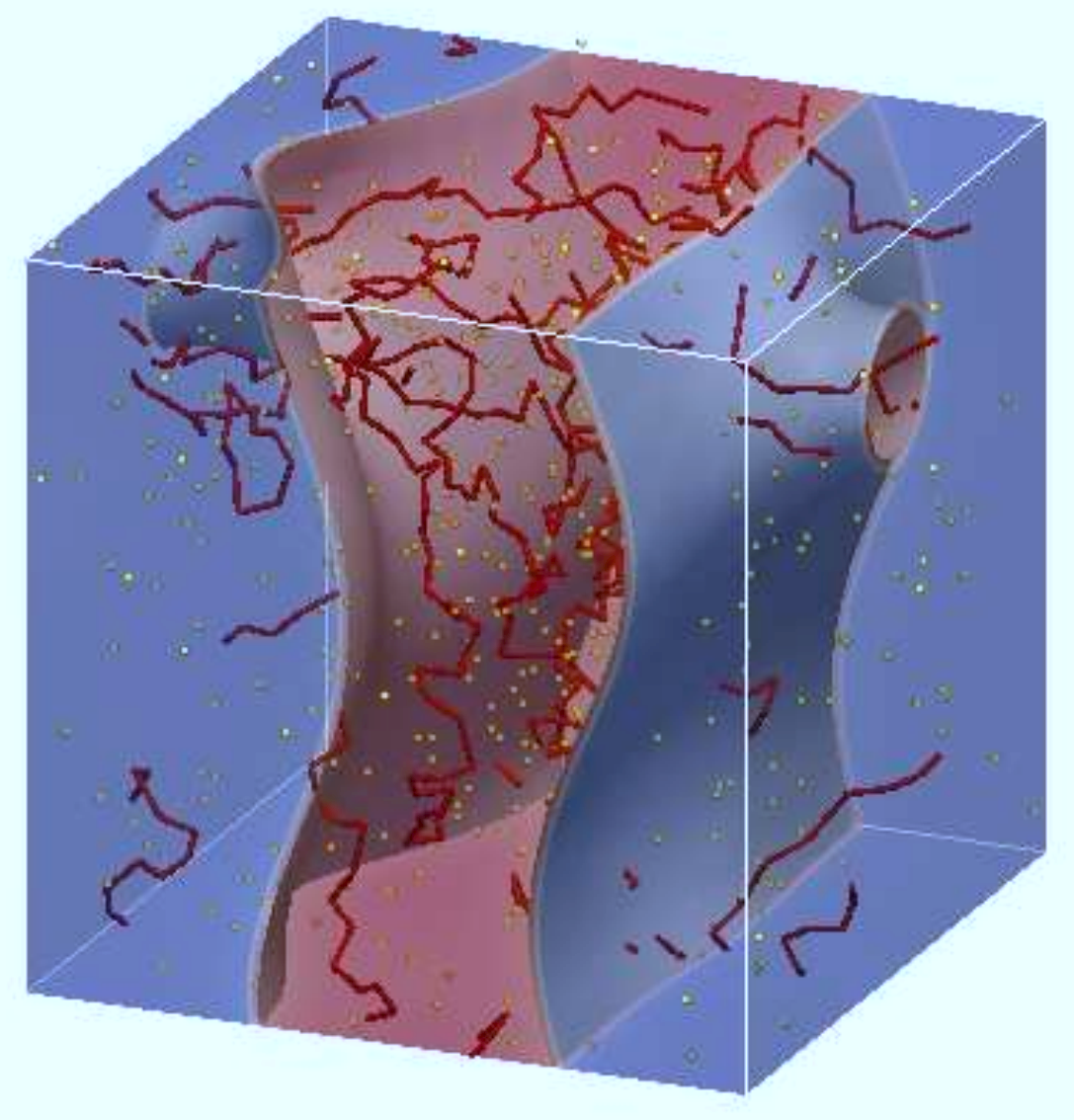}
\includegraphics[bb=142 180 500 542,clip,angle=90,width=0.32\columnwidth]{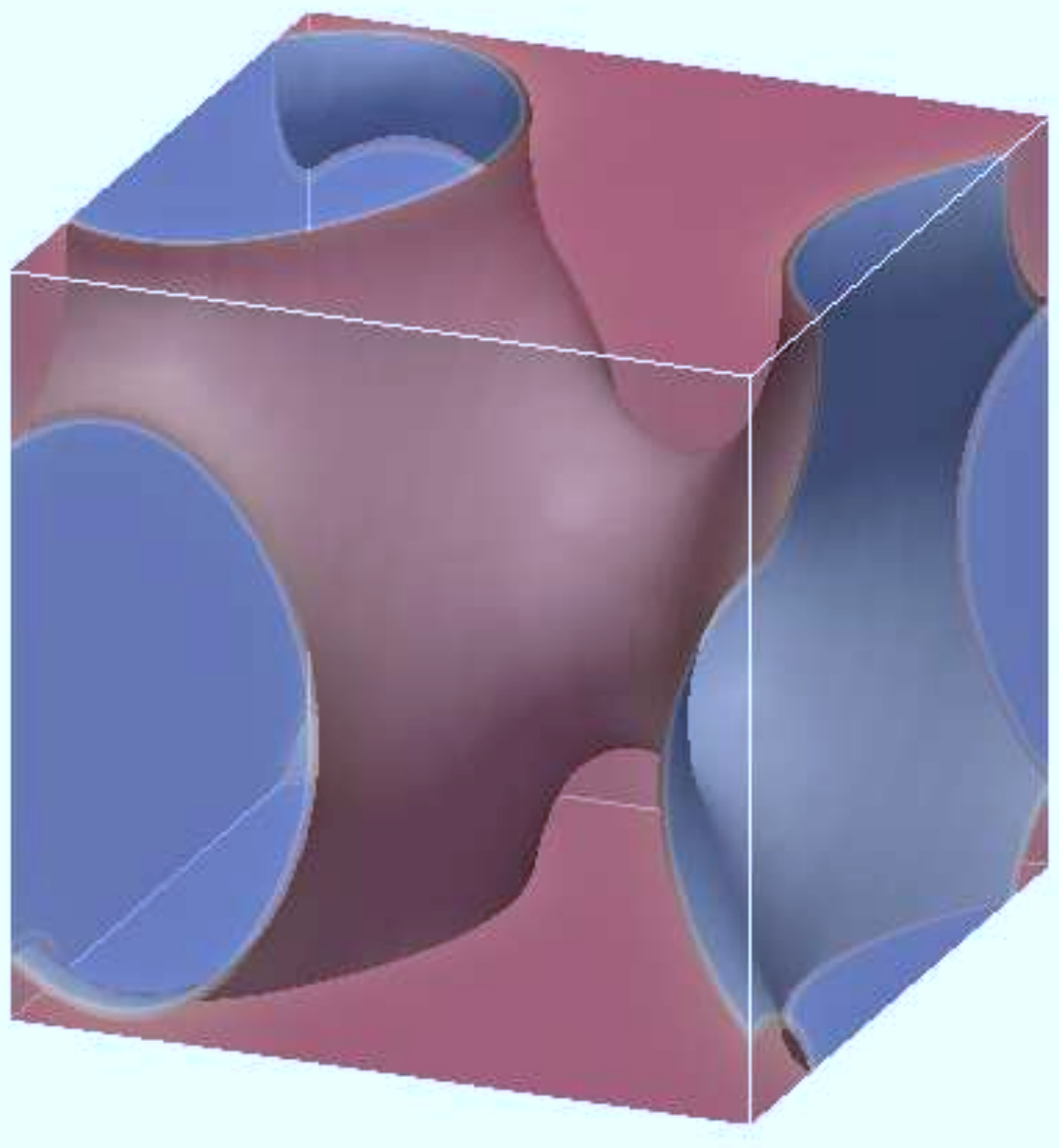}
\includegraphics[bb=142 180 500 542,clip,angle=90,width=0.32\columnwidth]{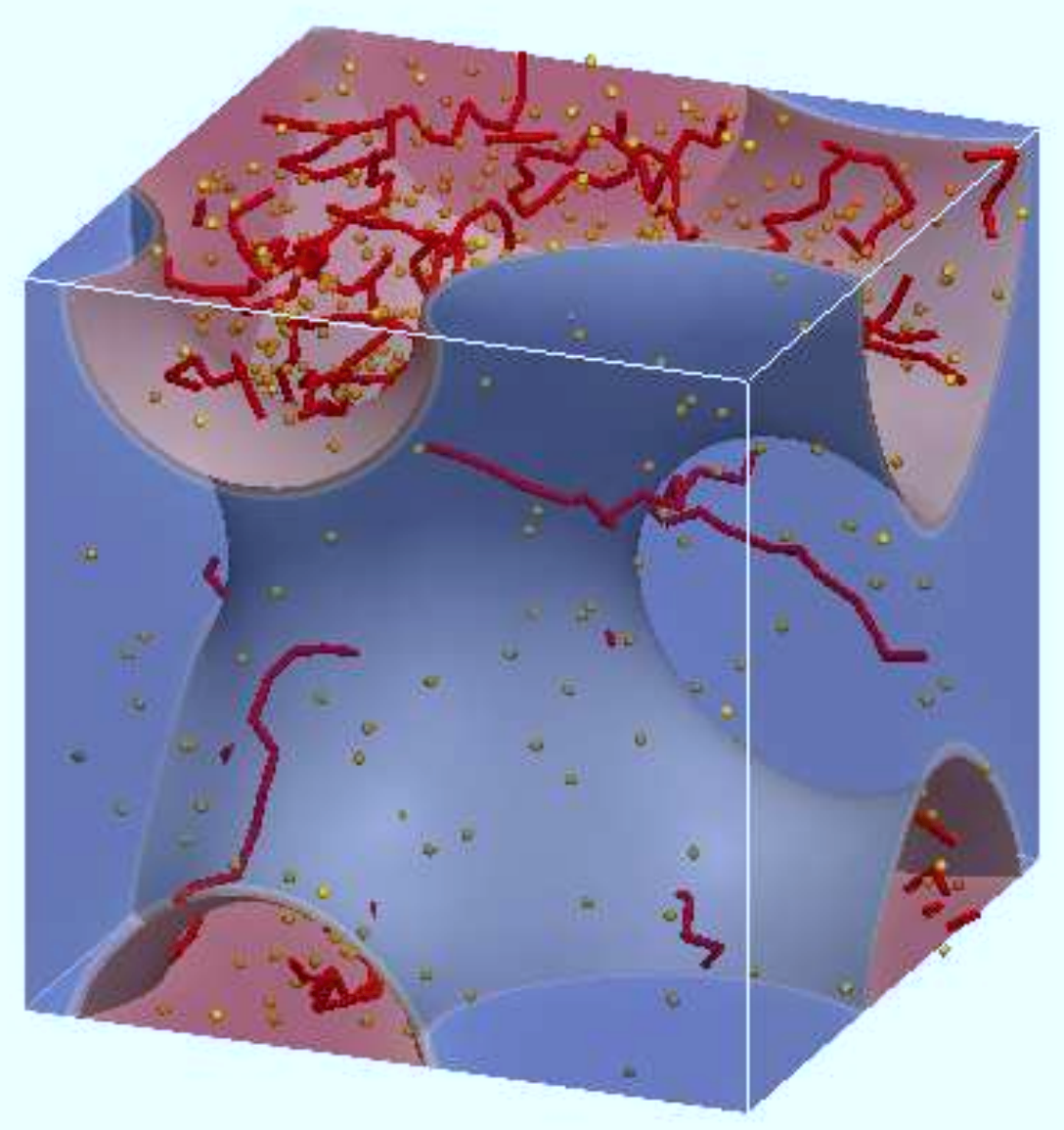}
\includegraphics[bb=142 180 500 542,clip,angle=90,width=0.32\columnwidth]{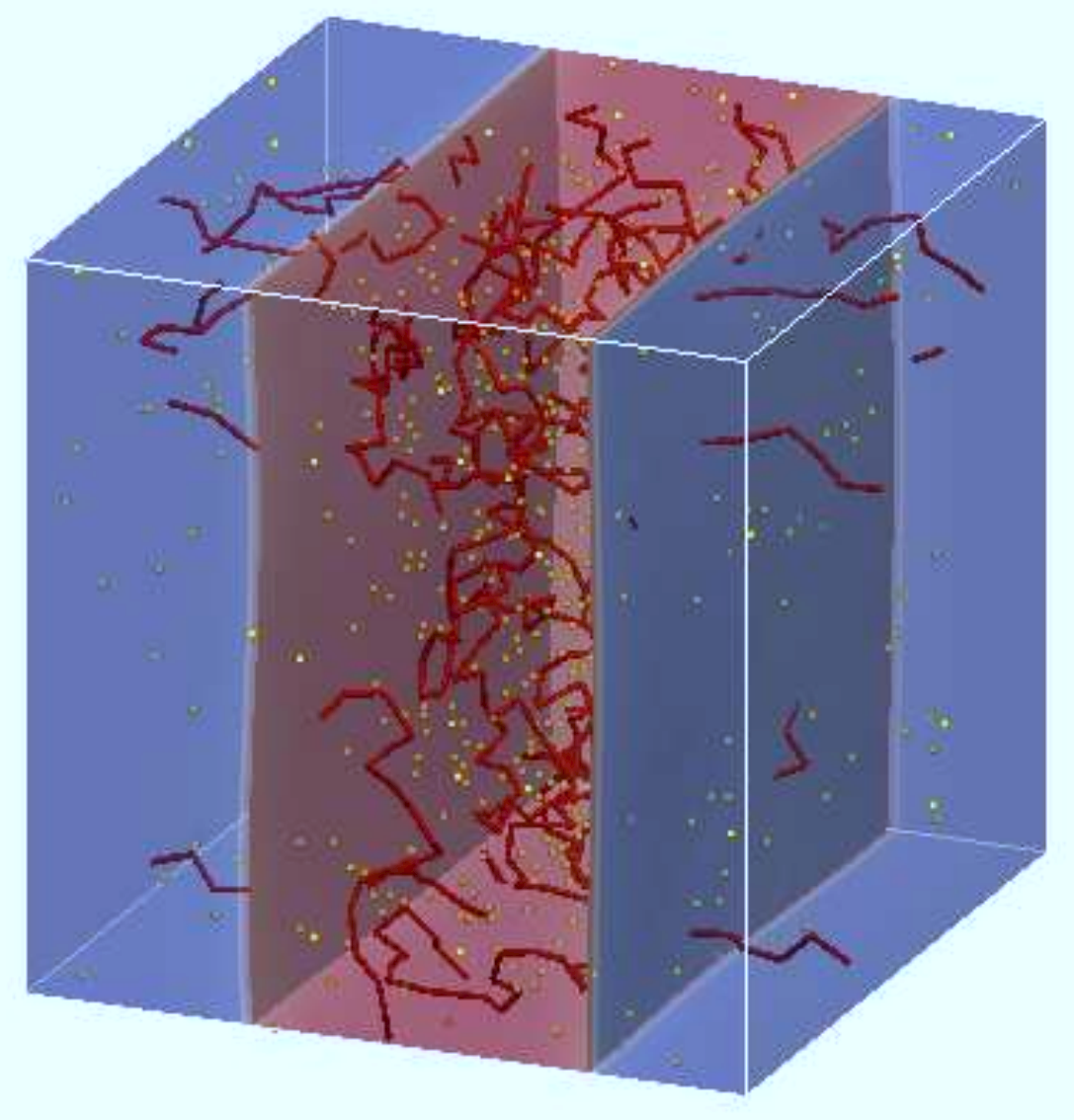}
\includegraphics[bb=142 180 500 542,clip,angle=90,width=0.32\columnwidth]{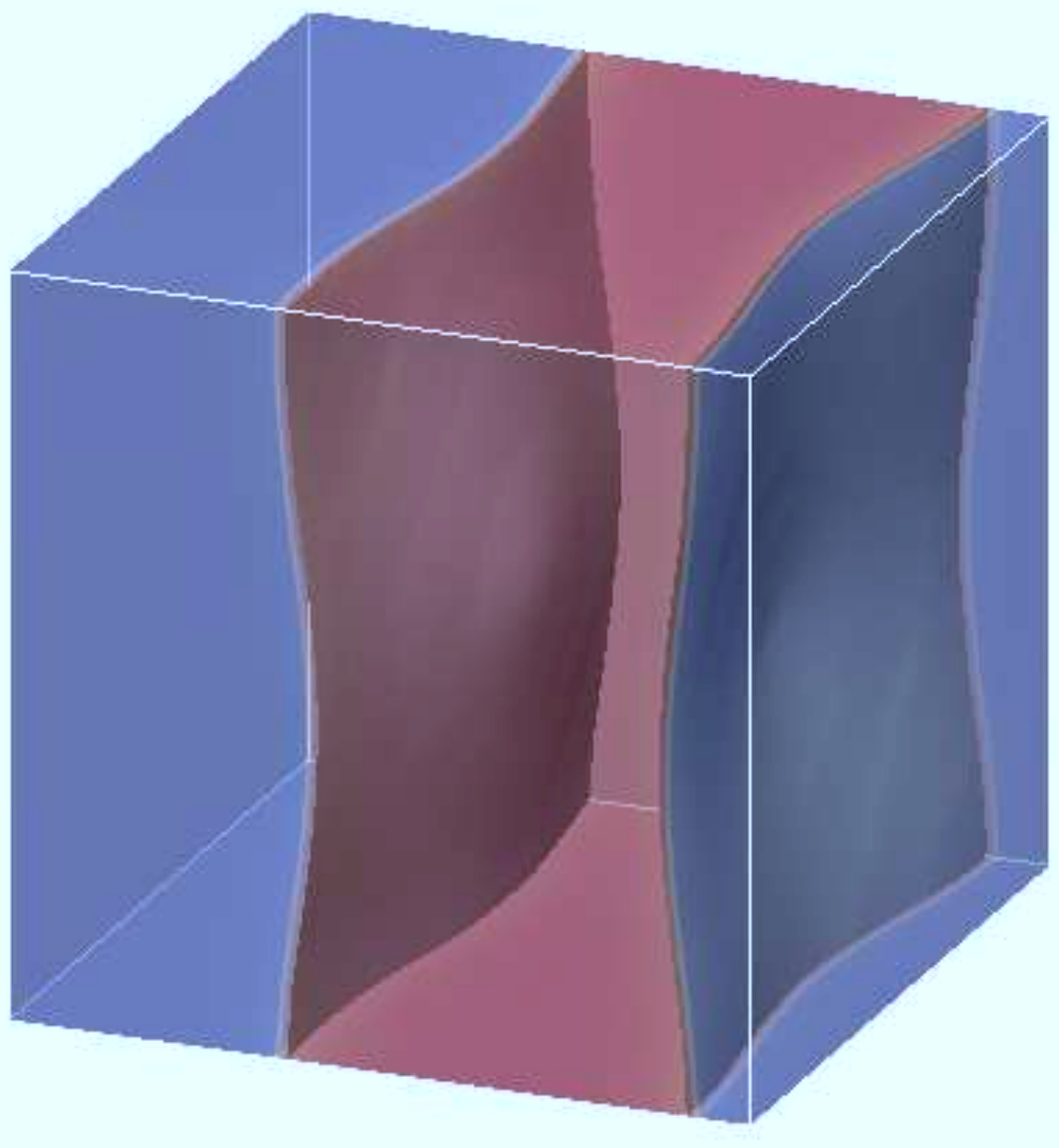}
\includegraphics[bb=142 180 500 542,clip,angle=90,width=0.32\columnwidth]{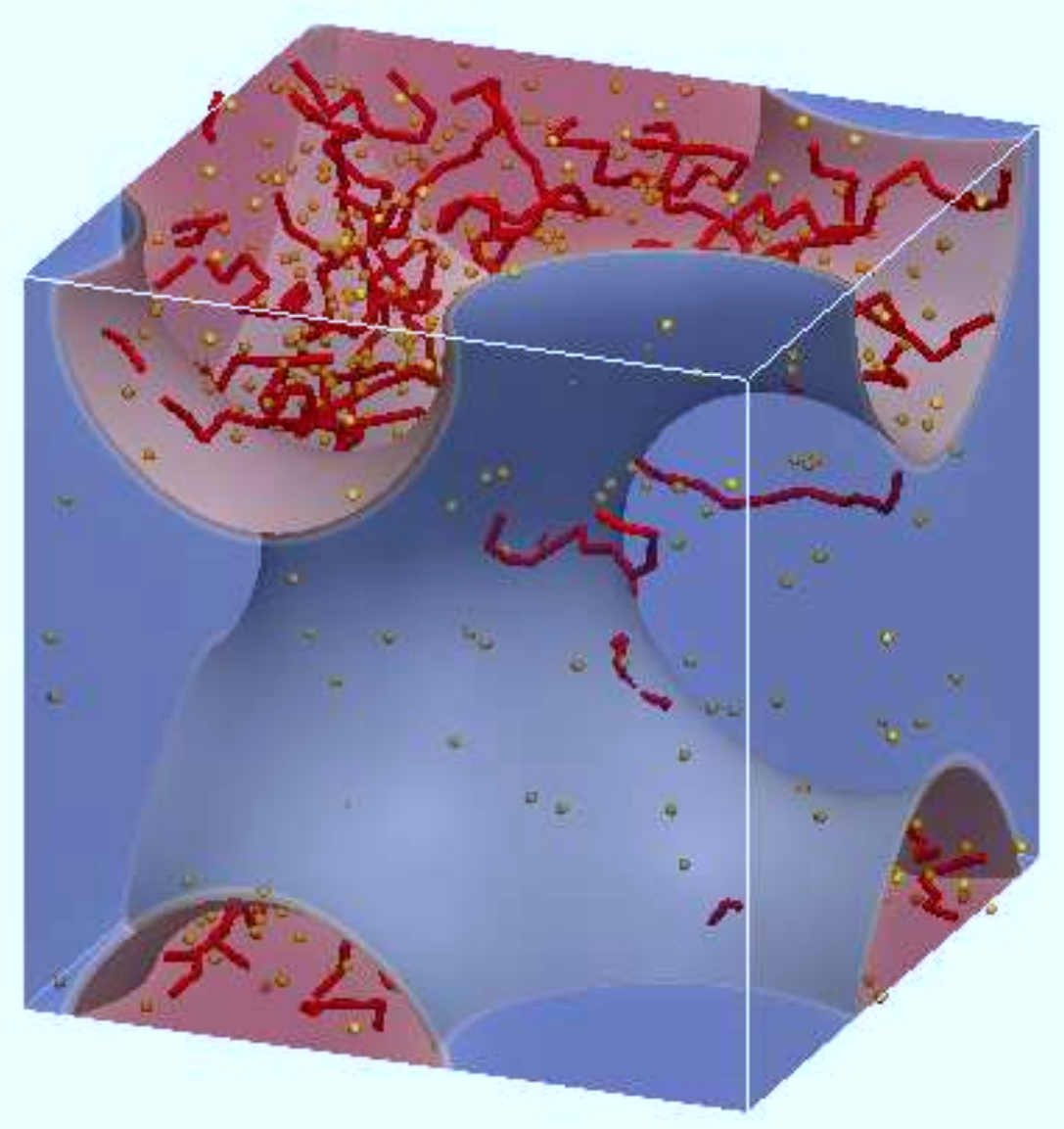}
\includegraphics[bb=142 180 500 542,clip,angle=90,width=0.32\columnwidth]{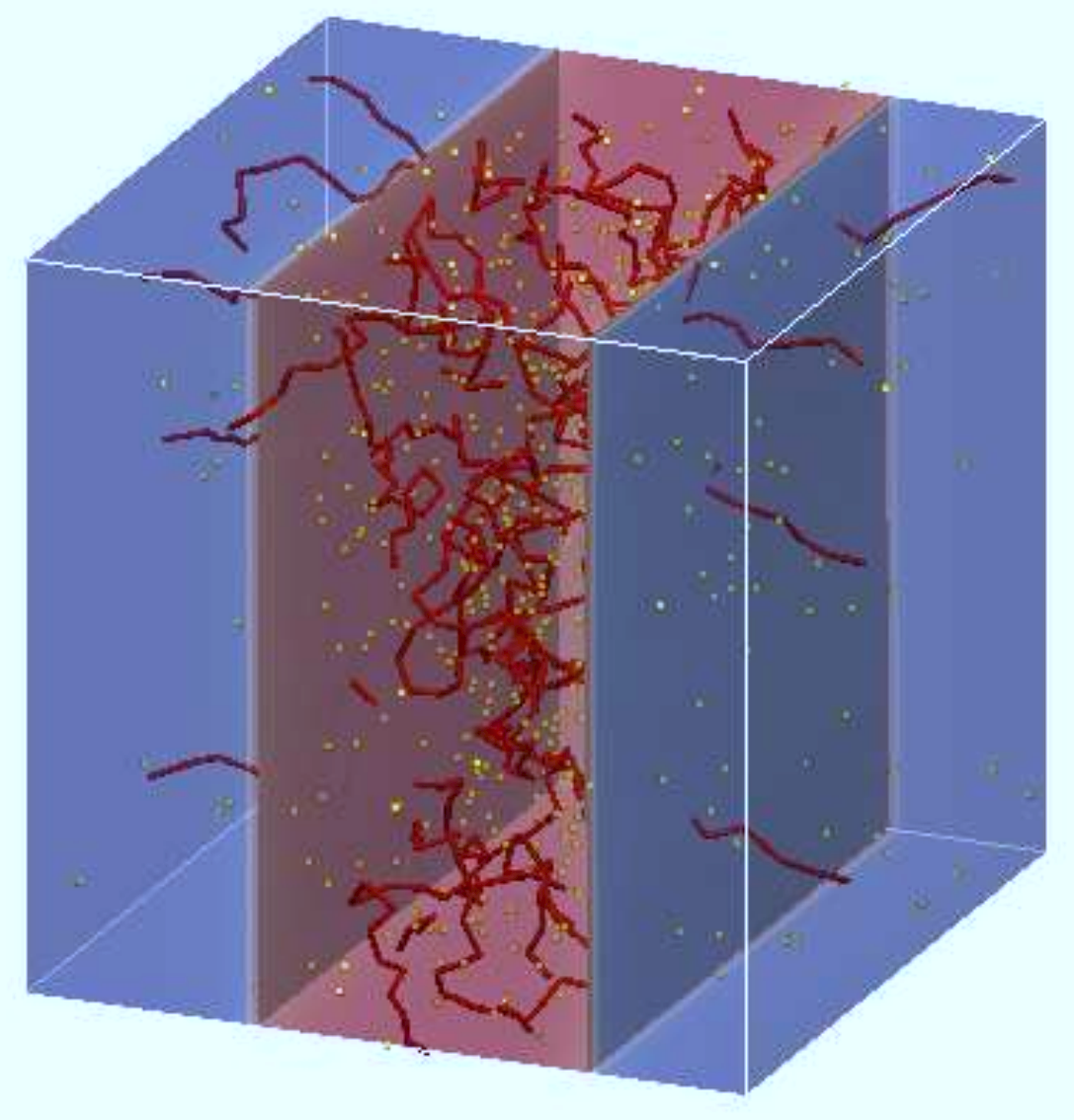}
\caption{Bicomponent fluid in absence (left) and presence of polyelectrolytes in solution (center: no added salt;  right: with implicit salt)\label{PEs}}
\end{figure}

In absence of particles, the two components separate macroscopically
(due to thermal fluctuations) at the end of a relatively long  domain
coarsening process, during which a bicontinuous phase is seen as a
metastable state (see Fig.~\ref{PEs}, left column). If polyelectrolytes
are added (central column) at random positions to the initial,
uniform fluid density, then parallel tubular structures are formed
at the beginning, to quickly evolve in a bicontinuous phase with
the polyelectrolyte mostly confined to the $A$ component. Only few
polymers are crossing the interface to the $B$ component, while
counterions, despite having the same coupling constants with the
fluid as the polyelectrolyte, can be found in considerable amount
also in the $B$ component. 
The reason for this behavior can be traced back to the fact that
the monomers are bonded with their neighbors along the chain,
therefore realizing a higher local density of interaction centers,
contrarily to the counterions, which are free to move apart. For
this reason, the thermal energy is enough to spread the counterions,
but not the polymers, through the $B$ component.
The bicontinuous phase remained stable
for the whole duration of the simulation, and is therefore either
an extremely long-lived metastable state, or, possibly, the stable
state of the system.

To see to what extent the electrostatic interaction contributes to the stabilization of the bicontinuous structure, we modeled the presence of added salt to the solution. 
In order to check the contribution of the electrostatic interaction only we replaced the Coulomb potential with a screened, Debye-H\"uckel one (instead of physically adding salt ions, which would have changed, e.g., also the entropy of the system),
\begin{equation}
U^{DH}_{ij}(r_{ij})=
\begin{cases}q_iq_j\ell_B\exp({-\kappa r_{ij}})/r_{ij} & r_{ij} < r_c\\
0 & r_{ij}\ge r_c
\end{cases},
\end{equation}
with screening length $k_D^{-1}=4\ell_B$ and cut-off $r_c=6\ell_B$, so that at the cut-off distance the screened potential of a ion pair is only about $0.04 k_BT$.  With the screened Coulomb interaction, similar tubular structures as for the unscreened case can be seen in the initial part of the simulation, but they do not evolve into a stable bicontinuous state, and collapse instead quickly into the macroscopic separated phase.  The formation process of the macroscopic separated phase is completed noticeably faster than in absence of polyelectrolytes, where the bicontinuous structure is a relatively long-lived metastable state.

\subsection{Quasi-2D ferrofluid emulsions.}
Ferrofluids are a class of superparamagnetic liquids composed of
ferromagnetic particles stabilized with surfactants and suspended
in a carrier fluid\cite{rosensweig97}. Magnetic particles in ferrofluids are usually of the 
size of few nanometers, and are therefore suspended thanks to
Brownian motion. The latter is comparable in strength to the magnetic
dipolar interaction and makes ferrofluids a notable example of
composite, magnetic soft-matter. When a ferrofluid is added to an
immiscible fluid, a so-called ferrofluid emulsion is formed, showing
the appearance of ferrofluid droplets\cite{bibette93,liu95,zakinyan11}.
Other examples of magnetic emulsions include ternary systems of two
immiscible liquids stabilized by magnetic particles at the interface,
forming a magnetic Pickering emulsion\cite{kaiser09,brown12}.

Ferrofluid emulsions have a high potential in
microfluidics\cite{bremond08,thiam09}, analytical\cite{gijs04} and
optical\cite{philip12} applications.  For all these applications,
the deformation of droplets in dependence of the external applied
magnetic field is of primary importance, as the magnetic permeability
of the emulsion depends strongly on the droplet shape, due to the
demagnetizing field effects.  In weak fields, the shape of ferrofluids
droplets is quite close to an ellipsoid of revolution elongated
along the direction of the external magnetic
field\cite{bacri82a,afkhami10,ivanov12}. The degree of elongation
in weak fields is by now fully understood and is well described by
both the pressure-mechanical\cite{blums97} and energy-minimization
approaches\cite{bacri82b,ivanov12}, while the breakup process has
been studied using a Lattice-Boltzmann approach in the full continuum
approximation, i.e., using consitutive equations to represent the
response of the fluid to the magnetic field\cite{falcucci09,falcucci10}.

\begin{figure}
\begin{center}
\includegraphics[bb=125 220 410 582,clip,angle=270,width=0.75\columnwidth]{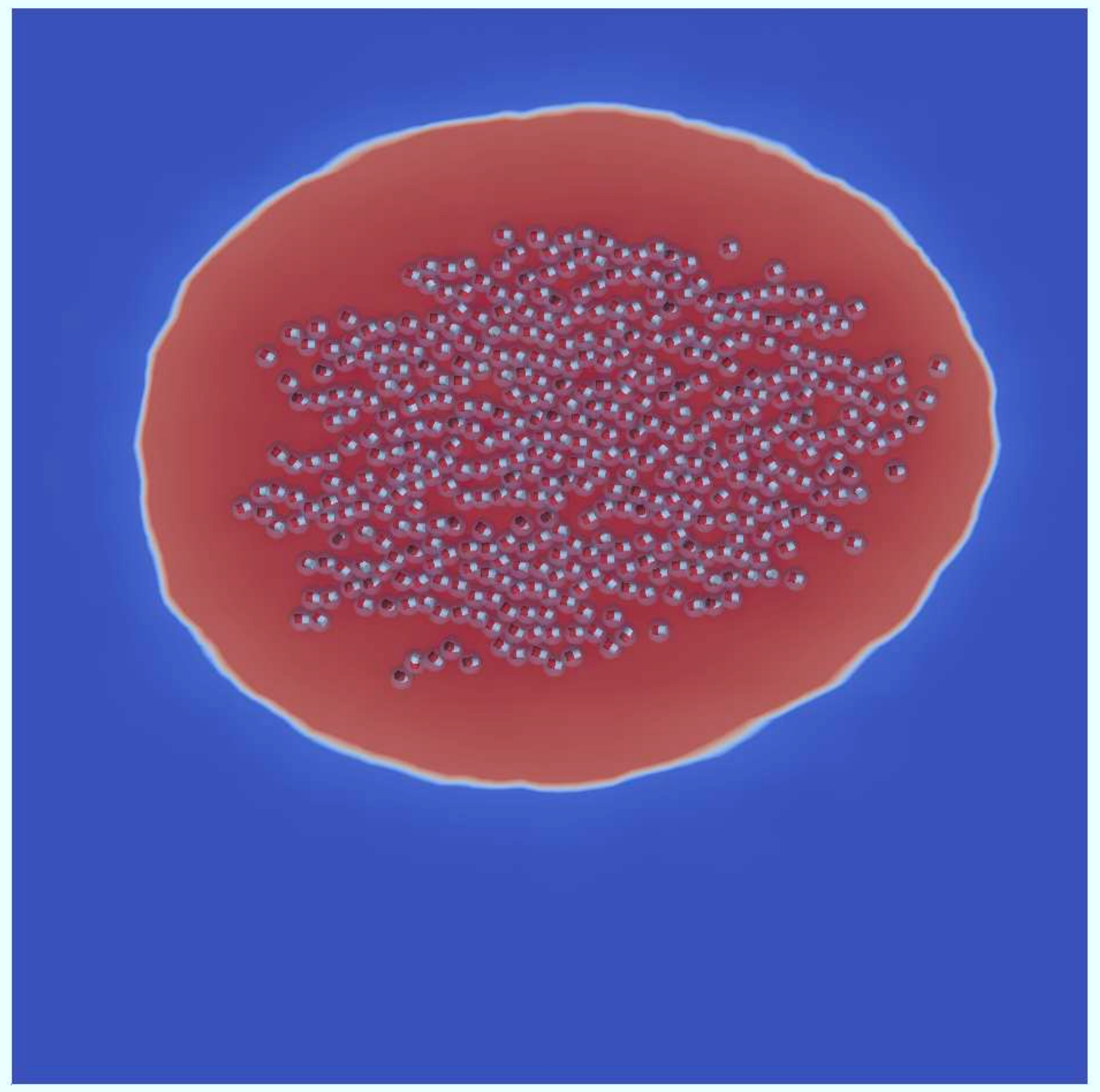}
\includegraphics[bb=100 40 500 722, clip,angle=270,width=0.75\columnwidth]{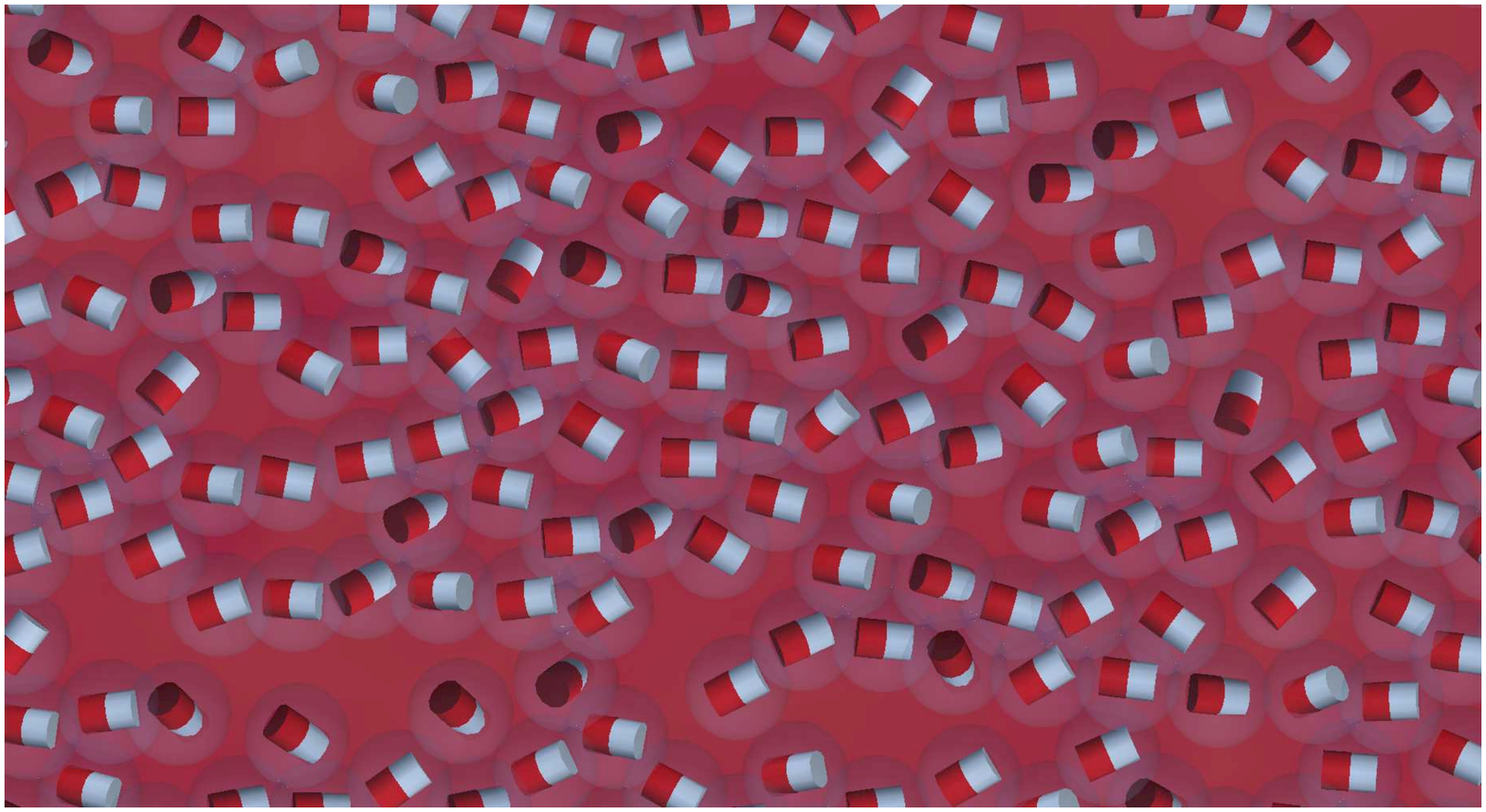}
\end{center}
\caption{Upper panel: quasi-2D ferrofluid emulsion under external magnetic field. Lower panel: detail of the central region of the drop.\label{ferrofluids}}
\end{figure}

Here we apply the coupling of particles to the Shan-Chen fluid to
show how it is possible to study ferrofluid emulsions under the
effect of an external magnetic field from a more microscopic point
of view, as an example of a complex bicomponent fluid/particle
mixture out of equilibrium. In this case, there is no need to
introduce constitutive equations, since the magnetic colloids
composing the ferrofluid are represented explicitly.
The system in analysis is a quasi-2D droplet simulated on a
$80\times80\times2$ lattice of spacing $a=1$, with the two fluid
components $A$ and $B$ having both average density 78.4 and with
Shan-Chen coupling parameters $g_{AB}=g_{BA}=0.0214$. In the droplet,
ferrofluids model particles have been placed, each of them being
represented using an excluded volume interaction (a WCA potential
with $\sigma=a$ and $\epsilon=k_BT$) that mimics the stabilizing
effect of the surfactant layer, and by the presence of a magnetic
point dipole at its center, free to rotate in all three spatial
directions and interacting via full 3D dipolar magnetic interaction,
whereas their centers are fixed in the droplet plane.

The system has been simulated at constant temperature.  Rigid body
equations of motion are integrated by taking into account in this
case all forces and torques originating from the WCA and magnetostatic
potential. The magnetic interaction is computed by summing over all
pairs and using the minimum image convention prior calculating
distances. The simulation box has been chosen to be larger than
twice the size of the droplet (simulation snapshots in
Fig.~\ref{ferrofluids} show only details of the simulation box) to
avoid magnetic self-interaction of the droplet.  The ferrofluid
particles in monolayer, at the density and dipolar interaction
strength employed in the present work, are forming, in absence of
external field, short chains\cite{klokkenberg06,kantorovich08}.
Upon application of an external magnetic field, the chains are all
orienting in the direction of the field (see lower panel of
Fig.~\ref{ferrofluids}). The droplet, from the initial circular
shape (not shown), becomes prolate (upper panel of Fig.~\ref{ferrofluids}),
displaying also rather pronounced fluctuations of the surface.

\section{Conclusions}
In this paper we have presented a novel way to address problems in complex liquid-liquid interfaces, that allows to describe the dynamics of bicomponent fluids in presence of complex solutes and is particularly suited to describe systems where the thermal fluctuations are playing a dominant role.  The method is a combination of the ``Shan-Chen'' multicomponent \cite{SC,SC2,SC5,ShanDoolen,ShanDoolen2} variant of Lattice Boltzmann with off-lattice molecular dynamics, inspired by the coupling introduced by Ahlrichs and D\"unweg \cite{ahlrichs98,ahlrichs99,ahlrichs01} for homogeneous fluids. The generalization to bicomponent fluids has brought under new light the nature of the coupling, showing the presence of a deeper symmetry between the fluid and the particles that allows to treat them, to some extent, interchangeably. The method has been shown to be able to  model, \emph{inter alia}, the contact angle even for point-like particles,  the interfacial deformations when colloidal particles are crossing the dividing surface between two components, and the surfactant effect of added amphiphilic molecules.

The particle-fluid coupling has been implemented in the ESPResSo \cite{limbach2006espresso,arnold2013espresso} simulation package, and the availability of a broad range of particle interaction potentials has allowed us to model quickly a number of problems.

As a first example of a complex solute, we have simulated flexible polyelectrolytes with explicit counterions in a binary fluid. The introduction of polyelectrolytes in the fluid mixture that, otherwise, would separates macroscopically, induces the formation of bicontinuous structures.  The stability of these mesoscopic structures proved to be sensitive to the ionic strength of the solution, as with the introduction of salt the mixture starts separating again, showing that the strength of the electrostatic interaction regulates the emergence of the bicontinuous phase. Obtaining the phase diagram for such kind of emulsions as a function of the polymer content and ionic strength is an objective of our future investigations.

The second problem investigated is the effect of an external magnetic
field on the structure of a single (quasi-2D) ferrofluid emulsion droplet. 
Particles in the ferrofluid were simulated as point magnetic dipoles with
an excluded volume interaction. The formation of chains and their
preferential orientation under the effect of the external magnetic
field create an anisotropic local environment that induces the
change in geometrical shape of the droplet from circular to elongated,
as seen in experiments and predicted by analytical calculations. A
simulation of the droplet shape deformation with explicit ferrofluid
particles, to the best of our knowledge, has been never performed
before, and future extensions to three-dimensional droplets will
allow us investigating the behavior of ferrofluid emulsions in the
very high field regime, which is out of reach for actual theoretical
analysis.

\paragraph*{Acknowledgements.}\quad{}M. Sega and M. Sbragaglia kindly acknowledge funding from the European Research Council under the Europeans Community's Seventh Framework Programme (FP7/2007-2013) / ERC Grant Agreement no[279004].  M. Sega acknowledges support from FP7 IEF p.n. 331932 SIDIS.  S.S.K.~is grateful to RFBR grants mol-a 1202- 31-374 and mol-a-ved 12-02-33106, has been supported by Ministry of Science and Education of RF 2.609.2011 and by Austrian Science Fund (FWF): START-Projekt Y 627- N27. A.O.I.~acknowledges RFBR grant 13-01-96032\_Ural.

\section*{Appendix: Lattice Boltzmann details and the Chapman--Enskog expansion}

In this Appendix we give the details for the lattice Boltzmann algorithm used in the numerical simulations and provide details of the Chapman-Enskog analysis to characterize the hydrodynamic equations of motion. As for the Chapman-Enskog analysis, our goal is to determine the correct form of the forcing source term $\Delta^{g}_{\zeta i}$ that enables us to recover the advection diffusion Eq. (\ref{eq:cont}).  
\subsection*{Lattice Boltzmann Scheme.}
The LB equation used in the numerical simulations is
\begin{equation}\label{EQ:LBapp}
f_{\zeta i} ({\bm{r}} + \tau {\bm{c}} , t + \tau ) = f^{*}_{\zeta i} ({\bm{r}},t)= f_{\zeta i} ({\bm{r}},t) + \Delta_{\zeta i} +\Delta^{g}_{\zeta i}+\hat{\Delta}_{\zeta i}
\end{equation}
with the collisional operator given by
\be\label{eq:collis}
\Delta_{\zeta i}=\sum_{j} {\cal L}_{ij}(f_{\zeta j}-f^{(eq)}_{\zeta j})
\ee
where the expression for the equilibrium distribution is a result of the projection onto the lower order Hermite polynomials and the weights $w_i$ are {\it a priori} known through the choice of the quadrature
\be\label{feq}
f_{\zeta i}^{(eq)}=w_i \rho_{\zeta} \left[1+\frac{{\bm{u}} \cdot {\bm{c}}_i}{c_s^2}+\frac{{\bm{u}}{\bm{u}}:({\bm{c}}_i{\bm{c}}_i-\id)}{2 c_s^4} \right]
\ee
\begin{equation}\label{weights}
w_i=
\begin{cases}
1/3 & i=0\\
1/18 & i=1\ldots6\\
1/36 & i=7\ldots18
\end{cases},
\end{equation}
where $c_s$ is the isothermal speed of sound and ${\bm{u}}$ is the velocity to be determined with the Chapman-Enskog procedure. Note that constructing equilibrium distribution functions with the same (baricentric) velocities leads to the correct hydrodynamic equations as soon as the relaxation matrix is the same for all the components. Our implementation features a D3Q19 model with 19 velocities
\begin{equation}\label{velo}
{\bm{c}}_i=
\begin{cases}
(0,0,0) & i=0\\
(\pm 1,0,0), (0,\pm 1,0), (0,0,\pm 1) & i=1\ldots6\\
(\pm 1,\pm 1,0), (\pm 1,0,\pm 1), (0,\pm 1,\pm 1)  & i=7\ldots18
\end{cases}.
\end{equation}
that, with the weights Eq. (\ref{weights}), produces the following tensorial identities
\be\label{isotropy}
\sum_{i} w_i c_{i \alpha}=0 ; \hspace{.1in} \sum_{i} w_i c_{i \alpha} c_{i \beta}=c_s^2 \delta_{\alpha \beta}.
\ee
\be\label{isotropyb}
\sum_{i} w_i c_{i \alpha}c_{i \beta}c_{i \gamma}=0 ; \hspace{.1in} \sum_{i} w_i c_{i \alpha}c_{i \beta}c_{i \gamma} c_{i \zeta}=c_s^4 \Delta_{\alpha \beta \gamma \zeta}.
\ee
The operator ${\cal L}_{ij}$ in Eq. (\ref{eq:collis}) is the same for both components (this choice is appropriate when we  describe a symmetric binary mixture) and is constructed to have a diagonal representation in the so-called {\it mode space}:  the basis vectors ${\bm{e}}_i$ of mode space are constructed by orthogonalizing polynomials of the dimensionless velocity vectors ${\bm{c}}_i$ \cite{Dunweg,dunweg08,Dunweg2,DHumieres02}. The basis vectors are used to calculate a complete set of moments, the so-called modes $m_{\zeta k}=\sum_i {\bm{e}}_{ki} f_{\zeta i}$ ($k=0,...,18$). The lowest order modes are associated with the hydrodynamic variables. In particular, the zeroth order momenta give the densities for both components
\begin{equation}
\rho_{\zeta}=m_{\zeta 0}=\sum_{i} f_{\zeta i},  
\end{equation}
with the total density given by $\rho=\sum_{\zeta}m_{\zeta 0} =\sum_{\zeta}\rho_{\zeta}$. The next three momenta $\tilde{\bm{m}}_{\zeta}=(m_{\zeta 1}, m_{\zeta 2}, m_{\zeta 3})$, when properly summed over all the components, are related to the baricentric velocity of the mixture
\begin{equation}
{\bm{u}} \equiv \frac{1}{\rho}\sum_{\zeta} \tilde{\bm{m}}_{\zeta}  +\frac{1}{2 \rho}\tau {\bm{g}} = \frac{1}{\rho}\sum_{\zeta} \sum_i f_{\zeta i} {\bm{c}}_{i}+\frac{1}{2 \rho}\tau {\bm{g}}
\end{equation}
with the total force density given by ${g}_{\alpha}=\sum_{\zeta} {g}_{\zeta \alpha}$ (see below, Eq. (\ref{momentum_definition})).  The other modes are the bulk and the shear modes (associated with the viscous stress tensor), and four groups of kinetic modes which do not emerge at the hydrodynamical level \cite{Dunweg}. Since the operator ${\cal L}_{ij}$ is  diagonal in mode space, the collisional term describes a linear relaxation of the non-equilibrium modes
\be\label{MODES}
m^{*}_{\zeta k}=(1+\lambda_k)m_{\zeta k}+m_{\zeta k}^{g}+ \phi_k r_k
\ee
where the relaxation frequencies $-\lambda_k$ (i.e. the eigenvalues of $-{\cal L}_{ij}$) are related to the transport coefficients of the modes. The term $m_{\zeta k}^{g}$ is related to the $k$-th moment of the forcing source $\Delta_{\zeta i}^{g}$ associated with a forcing term with density ${\bm{g}}_{\zeta}({\bm{r}}, t)$.  While the forces have no effect on the mass density, they transfer an amount ${\bm{g}}_{\zeta}\tau$ of total momentum to the fluid in one time step. Thermal fluctuations are represented by the stochastic term, $\phi_k r_k$, where $r_k$ is a Gaussian random number with zero mean and unit variance, and  $\phi_k$ is the amplitude of the mode fluctuation \cite{Dunweg}. The stochastic terms for the momentum and shear modes (leading to $\hat{\xi}_{\zeta \alpha}$ and $\hat{\sigma}_{\alpha \beta}$ in the hydrodynamic limit) represent a random flux and random stress. When dealing with two components ($\zeta=A,B$), we choose the same random number with opposite sign for the two components, so that $\hat{\xi}_{A \alpha}=-\hat{\xi}_{B \alpha}$. This allows to recover exactly the continuity equation for the whole mixture, $\partial_t \rho+{\bm{\nabla}} \cdot (\rho {\bm{u}}) = 0$, while keeping the fluctuating part in the equation for the order parameter $\phi=\rho_A-\rho_B$.  In the hydrodynamic limit, the variance of the random flux and random stress are fixed by the fluctuation-dissipation theorem to be~\cite{Dunweg,dunweg08,Dunweg2,Landau}
\begin{equation}
\langle \hat{\xi}_{\zeta \alpha}({\bm{r}},t)\hat{\xi}_{\zeta \alpha}({\bm{r}}',t') \rangle=2 k_B T \mu \, \delta(t-t')\delta({\bm{r}}-{\bm{r}}'),
\end{equation}
and
\begin{equation}
\langle \hat{\sigma}_{\alpha \beta}({\bm{r}},t)\hat{\sigma}_{\gamma \delta}({\bm{r}}',t') \rangle=2 k_B T \eta_{\alpha \beta \gamma \delta} \delta(t-t')\delta({\bm{r}}-{\bm{r}}'),
\end{equation}
respectively, where $\mu$ is the mobility and $\eta_{\alpha \beta \gamma \delta}$ is the tensor of viscosities formed out of the isotropic tensor $\delta_{\alpha \beta}$, the shear viscosity, $\eta_s$, and bulk viscosity, $\eta_b$ \cite{thampi}
\begin{dmath}
\eta_{\alpha\beta\gamma\delta}=\eta_s (\delta_{\alpha\gamma} \delta_{\beta\delta} + \delta_{\alpha\delta} \delta_{\beta\gamma} ) + \left(\eta_b-\frac{2}{3}\eta_s\right)  \delta_{\alpha\beta} \delta_{\gamma\delta}.
\end{dmath}
For the sake of simplicity the same viscosities for the two fluid phases are assumed. The transport coefficients $\mu$, $\eta_b$, $\eta_s$ are related to the relaxation times of the momentum, shear and bulk modes in ${\cal L}_{ij}$ (see Eq. (\ref{TRANSPORTCOEFF})).
\subsection*{Champan-Enskog Analysis.}
We next proceed with the Chapman-Enskog analysis. For simplicity, we do not treat the thermal fluctuations $\hat{\Delta}_{\zeta i}$ of the LB equation (\ref{EQ:LBapp}). The latter, once properly formulated in mode space (see Eq. (\ref{MODES})), result in a stochastic flux and stochastic tensor as given in Eqs. (\ref{eq:NS}) and (\ref{eq:cont}). The starting equation is therefore
\begin{equation}\label{EQ:LBapp-simple}
f_{\zeta i} ({\bm{r}} + \tau {\bm{c}} , t + \tau ) = f^{*}_{\zeta i} ({\bm{r}},t)= f_{\zeta i} ({\bm{r}},t) + \Delta_{\zeta i} +\Delta^{g}_{\zeta i}.
\end{equation}
In order to analyze the dynamics on the hydrodynamic scales, we have to coarse-grain time and space. We introduce a small dimensionless scaling parameter $\epsilon$. A coarse-grained length scale is introduced by writing ${\bm{r}}_1=\epsilon {\bm{r}}$, which corresponds to measuring positions with a coarse-grained ruler. We further introduce the convective time scale $t_1$ and the diffusive time scale $t_2$ by $t_1 = \epsilon t$ and $t_2 = \epsilon^2 t$. The deterministic LB equation is then
\begin{dmath}\label{eq:LB}
f_{\zeta i} ({\bm{r}}_1 + \epsilon \tau {\bm{c}}_i , t_1 + \ep \tau, t_2 + \ep^2 \tau ) = f_{\zeta i} ({\bm{r}}_1,t_1,t_2)+  \Delta_{\zeta i} +\Delta^{g}_{\zeta i}.
\end{dmath}
The LB equation written in terms of the coarse-grained variables can therefore be Taylor-expanded. Up to order ${\cal O}(\ep^2)$, we get
\begin{dmath}\nonumber
f_{\zeta i} ({\bm{r}}_1 + \epsilon \tau {\bm{c}} , t_1 + \ep \tau, t_2 + \ep^2 \tau ) = f_{\zeta i} ({\bm{r}}_1,t_1,t_2)+  \ep \tau \left(\frac{\partial}{\de t_1}+{\bm{c}}_i \cdot \frac{\de}{\de {\bm{r}}_1}\right)f_{\zeta i}  ({\bm{r}}_1,t_1,t_2) 
+\ep^2 \tau \left[\frac{\partial}{\de t_2}+\frac{\tau}{2} \left(\frac{\partial}{\de t_1}+{\bm{c}}_i \cdot \frac{\de}{\de {\bm{r}}_1}\right)^2   \right] f_{\zeta i}  ({\bm{r}}_1,t_1,t_2).
\end{dmath}
Similarly to the space-time variables, also the LB populations and the collision operator are expanded in powers of the scaling parameter $\ep$
\begin{dgroup}
\begin{dmath}
f_{\zeta i} =f_{\zeta i} ^{(0)}+\ep f_{\zeta i} ^{(1)}+\ep^2 f_{\zeta i} ^{(2)}+{\cal O}(\ep^3) 
\end{dmath}
\begin{dmath}
\Delta_{\zeta i} =\Delta_{\zeta i} ^{(0)}+\ep \Delta_{\zeta i} ^{(1)}+\ep^2 \Delta_{\zeta i} ^{(2)}+{\cal O}(\ep^3) 
\end{dmath}
\begin{dmath}
\Delta_{\zeta i}^g=\ep \Delta_{\zeta i}^{g(1)}+\ep^2 \Delta_{\zeta i}^{g(2)}+{\cal O}(\ep^3).
\end{dmath}
\end{dgroup}
Since the conservation laws hold on all scales, the collision operator must satisfy mass and global momentum conservation at all orders, that means
\be\label{Constraint}
\sum_i \Delta_{\zeta i}^{(k)}=0 \hspace{.2in } \sum_{\zeta{}i}  \Delta_{\zeta i}^{(k)}{\bm{c}}_i=0 
\ee
for all $k$. Using these expansions in Eq. (\ref{eq:LB}) we find
\begin{dmath}
\label{CE}
\ep \tau \left(\frac{\partial}{\de t_1}+{\bm{c}}_i \cdot \frac{\de}{\de {\bm{r}}_1}\right)f^{(0)}_{\zeta i} +\ep^2 \tau \left[\frac{\partial}{\de t_2}+\frac{\tau}{2} \left(\frac{\partial}{\de t_1}+{\bm{c}}_i \cdot \frac{\de}{\de {\bm{r}}_1}\right)^2   \right] f^{(0)}_{\zeta i} +\ep^2 \tau  \left(\frac{\partial}{\de t_1}+{\bm{c}}_i \cdot \frac{\de}{\de {\bm{r}}_1}\right)f^{(1)}_{\zeta i}=\Delta_{\zeta i}^{(0)}+\ep \Delta_{\zeta i}^{(1)}+\ep^2 \Delta_{\zeta i}^{(2)} + \ep \Delta_{\zeta i}^{g(1)}+\ep^2 \Delta_{\zeta i}^{g(2)}
\end{dmath}
where we have neglected all terms of order ${\cal O}(\ep^3)$. The different orders in (\ref{CE}) can be treated separately and we get a hierarchy of equations at different powers of $\ep$
\begin{dgroup}\label{eq:CE}
\begin{dmath}
{\cal O}(\epsilon^0): {\Delta_{\zeta i}^{(0)}=0} 
\end{dmath}
\begin{dmath}
{\cal O}(\epsilon^1): \left(\frac{\partial}{\de t_1}+{\bm{c}}_i \cdot \frac{\de}{\de {\bm{r}}_1}\right)f^{(0)}_{\zeta i}=\frac{1}{\tau} \left(\Delta_{\zeta i}^{(1)}+\Delta_{\zeta i}^{g(1)}\right)
\end{dmath}
\begin{dmath}
{\cal O}(\epsilon^2): \left[\frac{\partial}{\de t_2}+\frac{\tau}{2} \left(\frac{\partial}{\de t_1}+{\bm{c}}_i \cdot \frac{\de}{\de {\bm{r}}_1}\right)^2   \right] f^{(0)}_{\zeta i} + \left(\frac{\partial}{\de t_1}+{\bm{c}}_i \cdot \frac{\de}{\de {\bm{r}}_1}\right)f^{(1)}_{\zeta i}=\frac{1}{\tau}\left(\Delta_{\zeta i}^{(2)}+\Delta_{\zeta i}^{g(2)}\right).
\end{dmath}
\end{dgroup}
Using the second equation in the third we can rewrite the hierarchy of Eqs. (\ref{eq:CE}) in an equivalent but more convenient form 
\begin{dgroup}
\begin{dmath}
{\cal O}(\epsilon^0): \hspace{.1in} {\Delta_{\zeta i}^{(0)}=0}
\end{dmath}
\begin{dmath}
{\cal O}(\epsilon^1):  \hspace{.1in}\left(\frac{\partial}{\de t_1}+{\bm{c}}_i \cdot \frac{\de}{\de {\bm{r}}_1}\right)f^{(0)}_{\zeta i}=\frac{1}{\tau} \left(\Delta_{\zeta i}^{(1)}+\Delta_{\zeta i}^{g(1)}\right)
\end{dmath}
\begin{dmath}
{\cal O}(\epsilon^2): \hspace{.1in} \frac{\partial}{\de t_2} f^{(0)}_{\zeta i} +\frac{1}{2}\left(\frac{\partial}{\de t_1}+{\bm{c}}_i \cdot \frac{\de}{\de {\bm{r}}_1}\right)(f^{*(1)}_{\zeta i}+f^{(1)}_{\zeta i})=\frac{1}{\tau}\left(\Delta_{\zeta i}^{(2)}+\Delta_{\zeta i}^{g(2)}\right)
\end{dmath}
\end{dgroup}
where we have written $f^{*(1)}_{\zeta i}=f^{(1)}_{\zeta i}+\Delta_{\zeta i}^{(1)}+\Delta_{\zeta i}^{g(1)}$ for the ${\cal O}(\ep)$ post-collisional population. Since the momentum before and after the collisional-forcing step differ, the hydrodynamic momentum density is not uniquely defined. Any value between the pre- and the post-collisional value could be used. Consequently, there is an ambiguity which value to use for calculating the equilibrium distribution $f_{\zeta i}^{(eq)}$.  Without an {\it a priori} definition, we use the Chapman-Enskog expansion to deduce an appropriate choice. For this purpose, we introduce the following notations to distinguish between the global momentum densities obtained from the different orders of the Chapman-Enskog expansion
\be
{\bm{j}}^{\prime}= \sum_{\zeta{}i} {\bm{c}}_i f_{\zeta i}={\bm{j}}^{(0)}+\ep {\bm{j}}^{(1)}
\ee
where
\be\label{111}
{\bm{j}}^{(0)}=\sum_{\zeta{}i} {\bm{c}}_i f^{(0)}_{\zeta i} \hspace{.2in} {\bm{j}}^{(1)}=\sum_{\zeta}\sum_i {\bm{c}}_i f^{(1)}_{\zeta i}.
\ee
Since momentum is not conserved, ${\bm{j}}^{(1)}$ is not necessarily equal to zero.\\
{\bf Zeroth Order:} Here we identify $f_{\zeta i}^{(0)}$ with the equilibrium distribution $f_{\zeta i}^{(eq)}$, where we plug in ${\bm{u}}^{(0)}={\bm{j}}^{(0)}/\rho_{\zeta}$ for the flow velocity. The velocity ${\bm{u}}^{(0)}$ will be determined to get compliance with the macroscopic equations of motion.\\
{\bf First Order:}
The first two moments for the $\zeta$-th component at ${\cal O}(\ep)$ are
\be\label{CONT2F}
\frac{\de}{\de t_1}\rho_{\zeta}+\frac{\de}{\de {\bm{r}}_1} \cdot (\rho_{\zeta} {\bm{u}}^{(0)})=0
\ee
\begin{dmath}\label{MOM2F}
\frac{\de (\rho_{\zeta}{\bm{u}}^{(0)})}{\de t_1}+\frac{\de}{\de {\bm{r}}_1} \cdot \left(\sum_i f_{\zeta i}^{(eq)}{\bm{c}}_i{\bm{c}}_i\right)=\frac{\de (\rho_{\zeta}{\bm{u}}^{(0)})}{\de t_1}+\frac{\de}{\de {\bm{r}}_1} \cdot \left(p_{\zeta} \id+\rho_{\zeta} {\bm{u}}^{(0)}{\bm{u}}^{(0)} \right)=\frac{1}{\tau}({\bm{j}}^{*(1)}_{\zeta}-{\bm{j}}^{(1)}_{\zeta})
\end{dmath}
where, again, we have written $f^{*(1)}_{\zeta i}=f^{(1)}_{\zeta i}+\Delta_{\zeta i}^{(1)}+\Delta_{\zeta i}^{g(1)}$ for the ${\cal O}(\ep)$ post-collisional population. In Eq. (\ref{MOM2F}) we have used  $p_{\zeta}=c_s^2 \rho_{\zeta}$ to indicate the partial pressure for the $\zeta$-th component, being $p=\sum_{\zeta} p_{\zeta}$ the total pressure. The equations for the total momentum and the total momentum flux are obtained by taking the first and second moments, summing over $\zeta$, and considering that the forces transfer an amount ${\bm{g}} \tau$ of total momentum to the fluid in one time step
\begin{dmath}\label{MOMENTUM}
\frac{\de}{\de t_1} {\bm{j}}^{(0)}+\frac{\de}{\de {\bm{r}}_1} \cdot {\bm{\Pi}}^{(0)}=\frac{1}{\tau} \sum_{\zeta{}i} {\bm{c}}_i \left(\Delta_{\zeta i}^{(1)}+\Delta_{\zeta i}^{g(1)}\right) = {\bm{g}}^{(1)}
\end{dmath}
\be
\frac{\de}{\de t_1} {\bm{\Pi}}^{(0)}+\frac{\de}{\de {\bm{r}}_1} \cdot {\bm{\Phi}}^{(0)}=\frac{1}{\tau}\left({\bm{\Pi}}^{* (1)} -{\bm{\Pi}}^{(1)} \right).
\ee
We can first evaluate $\frac{1}{\tau}({\bm{j}}^{*(1)}_{\zeta}-{\bm{j}}^{(1)}_{\zeta})$ in (\ref{MOM2F}) as
\begin{dmath}
({\bm{j}}^{*(1)}_{\zeta}-{\bm{j}}^{(1)}_{\zeta})=\tau \left[\frac{\de (\rho_{\zeta}{\bm{u}}^{(0)})}{\de t_1}+\frac{\de}{\de {\bm{r}}_1} \cdot \left(p_{\zeta} \id+\rho_{\zeta} {\bm{u}}^{(0)}{\bm{u}}^{(0)} \right)\right]=\tau\left(\rho_{\zeta} D_{t_1}{\bm{u}}^{(0)}+\frac{\de p_{\zeta}}{\de {\bm{r}}_1} \right)
\end{dmath}
and $D_{t_1}{\bm{u}}^{(0)}=\left(\frac{\de }{\de t_1}+{\bm{u}}^{(0)} \cdot \frac{\de }{\de {\bm{r}}_1}\right){\bm{u}}^{(0)}$ can be obtained from the inviscid forced Euler equation, $D_{t_1}{\bm{u}}^{(0)}=-\frac{1}{\rho} \frac{\de p}{\de {\bm{r}}_1}+\frac{1}{\rho}{\bm{g}}^{(1)}$, so that
\be
({\bm{j}}^{*(1)}_{\zeta}-{\bm{j}}^{(1)}_{\zeta})=-\tau \left(\frac{\rho_{\zeta}}{\rho} \frac{\de p}{\de {\bm{r}}_1}-\frac{\de p_{\zeta}}{\de {\bm{r}}_1}\right)+\tau \left(\frac{\rho_{\zeta}}{\rho}{\bm{g}}^{(1)} \right).
\ee
A second relation is obtained from the relaxation properties in terms of the modes
\begin{dmath}
{\bm{j}}^{*(1)}_{\zeta}=(1+\lambda_M) {\bm{j}}^{(1)}_{\zeta}+\sum_{i} {\bm{c}}_i \Delta_{\zeta i}^{g(1)},
\end{dmath}
which implies 
\begin{dmath}
({\bm{j}}^{*(1)}_{\zeta}-{\bm{j}}^{(1)}_{\zeta}) = \lambda_M {\bm{j}}^{(1)}_{\zeta}+\sum_{i} {\bm{c}}_i \Delta_{\zeta i}^{g(1)}
\end{dmath}
and therefore
\begin{dmath}\label{J1}
{\bm{j}}^{(1)}_{\zeta}=\frac{1}{\lambda_M}\left[-\tau \left(\frac{\rho_{\zeta}}{\rho} \frac{\de p}{\de {\bm{r}}_1}-\frac{\de p_{\zeta}}{\de {\bm{r}}_1}\right)+\tau \frac{\rho_{\zeta}}{\rho}{\bm{g}}^{(1)} \right]-\frac{1}{\lambda_M}\sum_{i} {\bm{c}}_i \Delta_{\zeta i}^{g(1)}.
\end{dmath}
We can then evaluate ${\bm{\Pi}}^{(0)}$ and ${\bm{\Phi}}^{(0)}$ . This yields a similar result as that obtained with a single component flow \cite{Dunweg}, but with additional terms due to the forcing contribution in the momentum flux
\begin{dmath}\label{first}
\left({\Pi}^{* (1)}_{\alpha \beta}-{\Pi}^{(1)}_{\alpha \beta}\right)=\rho c_s^2 \tau \left(\frac{\de}{\de r_{1 \alpha}}u^{(0)}_{\beta}+\frac{\de}{\de r_{1 \beta}}u^{(0)}_{\alpha} \right)+\tau(u_{\alpha}^{(0)}g_{\beta}^{(1)}+g_{\alpha}^{(1)}u_{\beta}^{(0)})+{\cal O}(u^3).
\end{dmath}
A second relation is again obtained from the relaxation of the modes
\begin{dmath}\label{second}
\left({\Pi}^{* (1)}_{\alpha \beta}-{\Pi}^{(1)}_{\alpha \beta}\right)=\lambda_s\overline{\Pi}^{(1)}_{\alpha \beta}+\frac{\lambda_b}{3} \Pi^{(1)}_{\gamma \gamma} \delta_{\alpha \beta}+   \sum_{\zeta{}i}\Delta_{\zeta i}^{g (1)} c_{i\alpha}c_{i\beta}.
\end{dmath}
Solving the coupled Eqs. (\ref{first}) and (\ref{second}) yields 
\begin{dmath}\label{new1}
\overline{\Pi}^{* (1)}_{\alpha \beta}+\overline{\Pi}^{(1)}_{\alpha \beta}=\frac{\rho c_s^2 \tau (2+\lambda_s)}{\lambda_s}\left(\overline{\frac{\de}{\de r_{1 \alpha}}u_{\beta}^{(0)}}+\overline{\frac{\de}{\de r_{1 \beta}}u_{\alpha}^{(0)}} \right)+\frac{\tau (2+\lambda_s)}{\lambda_s}(\overline{u_{\alpha}^{(0)} g_{\beta}^{(1)}}+\overline{g_{\alpha}^{(1)} u_{\beta}^{(0)}})-\frac{2}{\lambda_s} \sum_{\zeta{}i}  \Delta_{\zeta i}^{g(1)} \overline{c_{i \alpha} c_{i \beta}}
\end{dmath}
\begin{dmath}\label{new1b}
\Pi^{* (1)}_{\alpha \alpha}+\Pi^{(1)}_{\alpha \alpha}=\frac{2\rho c_s^2 \tau (2+\lambda_b)}{\lambda_b}\frac{\de}{\de r_{1 \alpha}} u_{\alpha}^{(0)}+\frac{2 \tau (2+\lambda_b)}{\lambda_b}u_{\alpha}^{(0)}g_{\alpha}^{(1)}-\frac{2}{\lambda_b} \sum_{\zeta{}i} \Delta_{\zeta i}^{g(1)} c_{i \alpha} c_{i \beta}.
\end{dmath}
The additional terms due to the forcing can be compensated if the second moment of the forcing source is made to satisfy
\begin{dgroup}
\label{CON2}
\begin{dmath}
\sum_{\zeta{}i}\Delta_i^{g(1)} \overline{c_{i \alpha} c_{i \beta}}=\frac{\tau (2+\lambda_s)}{2}(\overline{u_{\alpha}^{(0)} g_{\beta}^{(1)}}+\overline{g_{\alpha}^{(1)} u_{\beta}^{(0)}}) \hspace{.1in}
\end{dmath}
\begin{dmath}
 \hspace{.4in} \sum_{\zeta{}i}\Delta_i^{g(1)} c_{i \alpha} c_{i \alpha}=(2+{\lambda_b})\tau u_{\alpha}^{(0)}g_{\alpha}^{(1)}.
\end{dmath}
\end{dgroup}
{\bf Second Order:} Proceeding to the order ${\cal O}(\ep^2)$, we start from
\begin{equation*}
\frac{\partial}{\de t_2} f^{(0)}_{\zeta i} + \frac{1}{2}\left(\frac{\partial}{\de t_1}+{\bm{c}}_i \cdot \frac{\de}{\de {\bm{r}}_1}\right)(f^{*(1)}_{\zeta i}+f^{(1)}_{\zeta i} )=\frac{1}{\tau} \Delta_{\zeta i}^{(2)}
\end{equation*}
so that, by taking the zeroth moment for the $\zeta$-th component, plus the information that in the momentum space we are relaxing according to ${\bm{j}}^{*(1)}_{\zeta}=(1+\lambda_M) {\bm{j}}^{(1)}_{\zeta}+\sum_{i} {\bm{c}}_i \Delta_{\zeta i}^g$, we find
\be\label{2nddens}
\frac{\de}{\de t_2} \rho_{\zeta} + \frac{\de}{\de {\bm{r}}_1} \cdot \left( \frac{(2+\lambda_M)}{2} {\bm{j}}^{(1)}_{\zeta}+\frac{1}{2}\sum_{i} {\bm{c}}_i \Delta_{\zeta i}^{g(1)} \right)=0.
\ee
The term ${\bm{j}}^{(1)}_{\zeta}$ was evaluated in (\ref{J1}) and, upon substitution in (\ref{2nddens}) we find
\begin{align}
\frac{\de}{\de t_2} \rho_{\zeta} + \frac{\de}{\de {\bm{r}}_1}\cdot\left\{ \frac{(2+\lambda_M)}{2 \lambda_M}\left[ -\tau \left(\frac{\rho_{\zeta}}{\rho} \frac{\de p}{\de {\bm{r}}_1}-\frac{\de p_{\zeta}}{\de {\bm{r}}_1}\right)\right.\right. \nonumber \\
\left.\left.+ \tau\frac{\rho_{\zeta}}{\rho}{\bm{g}}_{\zeta}^{(1)} -\sum_{i} {\bm{c}}_i \Delta_{\zeta i}^{g(1)} \right]+\frac{1}{2}\sum_{i} {\bm{c}}_i \Delta_{\zeta i}^{g(1)} \right\}=0.
\end{align}
The condition for the forces to be compatible with the pressure diffusion is found from
\begin{dmath}
-\frac{(2+\lambda_M)}{2 \lambda_M}\sum_{i} {\bm{c}}_i \Delta_{\zeta i}^{g(1)}+\frac{1}{2}\sum_{i} {\bm{c}}_i \Delta_{\zeta i}^{g(1)}=-\frac{2+\lambda_M}{2 \lambda_M}\tau {\bm{g}}^{(1)}_{\zeta}
\end{dmath}
yielding a constraint for the first order moment of the forcing term
\be\label{CON1}
\sum_{i} {\bm{c}}_i \Delta_{\zeta i}^{g(1)}=\frac{2+\lambda_M}{2}\tau {\bm{g}}^{(1)}_{\zeta}
\ee
and the continuity equation becomes
\begin{align}\label{DIFFUSIONb}
\frac{\de}{\de t_2} \rho_{\zeta} + \frac{\de}{\de {\bm{r}}_1} \cdot \left\{ \frac{(2+\lambda_M)}{2 \lambda_M}\left[ \tau \left(-\frac{\rho_{\zeta}}{\rho} \frac{\de p}{\de {\bm{r}}_1}+\frac{\de p_{\zeta}}{\de {\bm{r}}_1}\right)\right.\right.\nonumber\\
\left.\left.+\tau \left(-\frac{\rho_{\zeta}}{\rho}{\bm{g}}^{(1)}+{\bm{g}}_{\zeta}^{(1)} \right) \right] \right\}=0.
\end{align}
The first order moment for the whole mixture delivers
\begin{dmath}\label{ABmix}
\frac{\de}{\de t_2} {\bm{j}}^{(0)}+\frac{\de}{\de t_1} \left({\bm{j}}^{(1)}+\frac{1}{2}\tau {\bm{g}}^{(1)} \right)+\frac{1}{2}\frac{\de}{\de {\bm{r}}_1} \cdot \left({\bm{\Pi}}^{* (1)}+{\bm{\Pi}}^{(1)} \right)={\bm{g}}^{(2)}.
\end{dmath}
Inserting the results (\ref{new1}) and (\ref{new1b}) for ${\bm{\Pi}}^{(1)}$  in (\ref{ABmix}) gives
\begin{dmath}\label{MOMENTUMb}
\frac{\de}{\de t_2} j_{\alpha}^{(0)}+\frac{\de}{\de t_1} \left(j_{\alpha}^{(1)}+\frac{1}{2}\tau g_{\alpha}^{(1)} \right)+\rho c_s^2 \tau \frac{\de}{\de r_{1\beta}} \left[\frac{2+\lambda_s}{2\lambda_s}\left(\frac{\de}{\de r_{1 \alpha}}u_{\beta}^{(0)}+\frac{\de}{\de r_{1 \beta}}u_{\alpha}^{(0)}\right)+\frac{2+\lambda_b}{3 \lambda_b}\frac{\de}{\de r_{1 \gamma}} u_{\gamma}^{(0)}\delta_{\alpha \beta} \right] =g_{\alpha}^{(2)}.
\end{dmath}
After merging orders we arrive at the continuity equation for the species (using Eqs. (\ref{CONT2F}) and (\ref{DIFFUSIONb}))
\begin{dmath}\label{MERGING1}
\frac{\de}{\de t} \rho_{\zeta}+\frac{\de }{\de r_{\alpha}} (\rho_{\zeta} u_{\alpha}^{(0)}) \\=\mu \frac{\de}{\de r_{\alpha}}  \left[\left(\frac{\de p_{\zeta}}{\de r_{\alpha}}-\frac{\rho_{\zeta}}{\rho} \frac{\de p}{\de r_{\alpha}}\right)-\left(g_{\zeta \alpha}-\frac{\rho_{\zeta}}{\rho}g_{\alpha}\right) \right]
\end{dmath}
and the momentum equation for the mixture (using Eqs. (\ref{MOMENTUM}) and (\ref{MOMENTUMb}))
\begin{dmath}\label{MERGING1b}
\frac{\de}{\de t} \left(j_{\alpha}^{\prime}+\frac{1}{2}\tau g_{\alpha}\right)+\frac{\de}{\de r_{\beta}}\left(\rho c_s^2 \delta_{\alpha \beta}+\rho u^{(0)}_{\alpha}u^{(0)}_{\beta} \right)-\frac{\de}{\de r_{\beta}}\left[\eta_s\left(\frac{\de}{\de r_{\alpha}} u^{(0)}_{\beta}+\frac{\de}{\de r_{\beta}} u^{(0)}_{\alpha} -\frac{2}{3}\frac{\de}{\de r_{\gamma}} u^{(0)}_{\gamma} \delta_{\alpha \beta}\right)+\eta_b \frac{\de}{\de r_{\gamma}} u_{\gamma}^{(0)} \delta_{\alpha \beta} \right]=g_{\alpha}
\end{dmath}
where we have defined the following transport coefficients
\begin{dgroup}\label{TRANSPORTCOEFF}
\begin{dmath}
\mu=-\tau \left(\frac{1}{\lambda_M}+\frac{1}{2} \right)
\end{dmath}  
\begin{dmath}
\eta_s=-\rho c_s^2 \tau \left(\frac{1}{\lambda_s}+\frac{1}{2} \right) 
\end{dmath}  
\begin{dmath}
\eta_b=-\rho c_s^2 \tau \left(\frac{1}{\lambda_b}+\frac{1}{2} \right).
\end{dmath}
\end{dgroup}
Eqs. (\ref{MERGING1}) and (\ref{MERGING1b}) can be cast in the form of the Navier-Stokes equations (\ref{eq:NS}) and (\ref{eq:cont}) by using the following definition for the components of the diffusion current ${\bm{D}}$ 
\begin{dmath}
D_{\zeta \alpha}=\mu \left[\left(\frac{\de p_{\zeta}}{\de r_{\alpha}}-\frac{\rho_{\zeta}}{\rho} \frac{\de p}{\de r_{\alpha}}\right)-\left(g_{\zeta \alpha}-\frac{\rho_{\zeta}}{\rho}g_{\alpha}\right) \right],
\label{eq:comp_D}
\end{dmath}
of the viscous stress tensor ${\bm{\Pi}}$ 
\begin{dmath}
\Pi_{\alpha \beta} =\eta_s{\left(\frac{\de}{\de r_{\alpha}} u_{\beta}+\frac{\de}{\de r_{\beta}} u_{\alpha} -\frac{2}{3}\frac{\de}{\de r_{\gamma}} u_{\gamma} \delta_{\alpha \beta}\right)}+\eta_b \frac{\de}{\de r_{\gamma}} u_{\gamma} \delta_{\alpha \beta},
\label{eq:comp_Pi}
\end{dmath}
and of the total hydrodynamic momentum density (which is used in the equilibrium distribution):
\begin{equation}
{\bm{j}} \equiv {\bm{j}}^{(0)} \equiv {\bm{j}}^{\prime}+\frac{1}{2}\tau {\bm{g}}=\sum_{\zeta{}i} f_{\zeta i} {\bm{c}}_i+\frac{1}{2}\tau {\bm{g}}.
\label{momentum_definition}
\end{equation}
Note that this implies
\be
\sum_{\zeta{}i}  f_{\zeta i}^{(eq)} {\bm{c}}_i={\bm{j}} \mathrm{\quad{}and\quad{} } \sum_{\zeta}\sum_i f_{\zeta i}^{(neq)} {\bm{c}}_i=-\frac{1}{2}\tau {\bm{g}}.
\ee
The above definition corresponds to the arithmetic mean of the pre- and post-collisional global momentum density. The forcing term is determined from the conditions (\ref{CON1}) and (\ref{CON2}), and can be written as
\begin{align}
\Delta_{\zeta i}^{g}=\frac{w_i \tau}{c_s^2} \left(\frac{2+\lambda_M}{2}\right) {\bm{g}}_{\zeta} \cdot {\bm{c}}_i\\ +\frac{w_i \tau}{c_s^2} \left[\frac{1}{2c_s^2} {\bm{G}} : ({\bm{c}}_i {\bm{c}}_i-c_s^2 \id ) \right],
\end{align}
where the components of tensor $G$ are defined as 
\begin{dmath}
G_{\alpha \beta}=\frac{2+\lambda_s}{2}\left(u_{\alpha} g_{\beta}+g_{\alpha} u_{\beta}-\frac{2}{3} u_{\gamma} g_{\gamma} \delta_{\alpha \beta} \right)+\frac{2+\lambda_b}{3} u_{\gamma} g_{\gamma} \delta_{\alpha \beta}.
\end{dmath}

\footnotesize{
\providecommand*{\mcitethebibliography}{\thebibliography}
\csname @ifundefined\endcsname{endmcitethebibliography}
{\let\endmcitethebibliography\endthebibliography}{}

}
\end{document}